\documentclass{iopart}
\usepackage{iopams}
\usepackage{graphicx}
\usepackage{color}

\expandafter\let\csname equation*\endcsname\relax
\expandafter\let\csname endequation*\endcsname\relax
 \usepackage{amsmath}

\newcommand{\bra}[1]{\langle#1|}
\newcommand{\ket}[1]{|#1\rangle}
\newcommand{\vb}[1]{\pmb{#1}}

\usepackage{xifthen}
\newcommand{\vq}[2][]{                
  \ifthenelse{\isempty{#1}}           %
    { \hat{\pmb{#2}} }             
    { \hat{\pmb{#2}}_\mathrm{#1} } 
}
\newcommand{\tq}[2][]{                
  \ifthenelse{\isempty{#1}}           %
    { \mathbb{#2} }                   
    { \mathbb{#2}^\mathrm{#1} }       
}
\newcommand{\vs}[2][]{                
  \ifthenelse{\isempty{#1}}           %
    { \mathbf{#2} }                   
    { \mathbf{#2}^\mathrm{#1} }       
}

\newcommand{\SD}[1]{{#1}}

\begin{document}

\title[Quantum noise of non-ideal
Sagnac speed meter interferometer with asymmetries]{Quantum noise of non-ideal
Sagnac speed meter interferometer with asymmetries}

\author{
S~L~Danilishin$^{1}$, C~Gr\"af$^{2}$, S~S~Leavey$^{2}$, J~Hennig$^{2}$, 
E~A~Houston$^{2}$, D~Pascucci$^{2}$, S~Steinlechner$^{2}$, J~Wright$^{2}$ 
and S Hild$^{2}$
}
\ead{Stefan.Danilishin@ligo.org}
\vskip 1mm
\address{$^{1}$\,School of Physics, University of Western Australia, 35 
Stirling Hwy, Crawley 6009, Australia}
\address{$^{2}$\,SUPA, School of Physics and Astronomy, The University 
of Glasgow, Glasgow, G12\,8QQ, UK}

\begin{abstract}
The speed meter concept has been identified as a technique that can
potentially provide laser-interferometric measurements at 
a sensitivity level which surpasses the Standard Quantum 
Limit (SQL) over a broad frequency range. As with other 
sub-SQL measurement techniques, losses play a central role
in speed meter interferometers and they ultimately determine
the quantum noise limited sensitivity that can be achieved.
So far in the literature, the quantum noise limited sensitivity has
only been derived  for lossless or lossy cases using certain
approximations (for instance that the arm cavity round trip
loss is small compared to the arm cavity mirror transmission).
In this article we present a generalised, analytical treatment
of losses in speed meters that allows accurate calculation
of the quantum noise limited sensitivity of Sagnac speed meters
with arm cavities. In addition, our analysis allows us to take
into account potential imperfections in the interferometer
such as an asymmetric beam splitter or differences of the 
reflectivities of the two arm cavity input mirrors. Finally,
we use the examples of the proof-of-concept Sagnac speed meter
currently under construction in Glasgow and a potential
implementation of a Sagnac speed meter in the Einstein Telescope
(ET) to illustrate how our findings affect Sagnac speed meters
with meter- and kilometre-long baselines.
\end{abstract}


\section{Introduction}
\label{sec:intro}
The sensitivity of state-of-the-art laser-interferometric
gravitational wave detectors,
such as the Advanced LIGO detector \cite{Harry10} currently being
commissioned, will be limited over most frequencies in 
its detection band by so-called \emph{quantum noise}. Quantum 
noise comprises of two components: sensing noise (photon shot noise)
at high frequencies and back-action noise (photon radiation
pressure noise) at low frequencies. One strategy for significant 
quantum noise reduction is to replace conventional position
meters in these interferometers with a speed meter \cite{Braginsky90}.
This allows, at least partially, the evasion of back-action
noise and therefore provides the possibility of broadband 
sub-SQL measurements \cite{Liv.Rv.Rel.15.2012}.
 
The first implementation of a laser-interferometric speed meter 
was based on a Michelson interferometer employing an additional
sloshing cavity in its output port \cite{00a1BrGoKhTh,Purdue01,Purdue02}. In 2003, it
was then shown by Chen that a Sagnac interferometer has inherent 
speed meter characteristics \cite{Chen2002}. This article 
also included the first analytical treatment of the achievable suppression
of back-action noise in a Sagnac speed meter, but did 
not include treatment of any effects arising from optical losses. Although the loss analysis in Michelson-based 
sloshing speed meters was done in \cite{Purdue02}, the first treatment of loss
for a Sagnac speed meter was undertaken by Danilishin \cite{Danilishin04}.
In the same article a new concept for a realisation 
of a Sagnac speed meter based on polarisation optics was suggested.

In the context of the Einstein Telescope design
 \cite{et_punturo2010, Hild2011},
the analytical analysis of losses in speed meter interferometers
was extended to Sagnac interferometers employing arm cavities
as well as recycling techniques \cite{phd.MuellerEbhardt} and it 
was shown using theoretical analyses that speed meter 
interferometers can significantly outperform traditional 
Michelson interferometers in terms of quantum noise
 \cite{ET-010-09,GRG.43.2.671_2011_Chen}. Additional work has shown that it is 
possible to implement a DC-readout technique \cite{Ward08, Hild09b} based on polarisation Sagnac interferometers \cite{Wang13}. 
Recently, the potential benefit of Sagnac speed meters for 
Advanced LIGO upgrades has been analysed and has also shown to be 
significant \cite{LIGO-T1200042-v1, Miao14}.
   
While there has been significant effort over the past 10 years 
to study aspects of speed meter configurations from a theoretical
point of view, so far the performance of the speed meter concept 
has not been demonstrated in an experiment. Therefore, we recently started
to set up a Sagnac speed meter proof-of-concept experiment, that
aims to demonstrate the reduction of back-action noise provided
by the speed meter \cite{Graef14}.

In this article we further advance the quantum noise models 
for Sagnac speed meters, firstly by including treatment for
asymmetries in the interferometer (such as an 
asymmetric beam splitter or arm cavity input-coupling mirrors
with different reflectivities), and secondly by providing 
a more general treatment of losses. Furthermore, the losses
do not rely on certain approximations, such as that arm cavity
losses are much smaller than the input mirror transmission, an
approximation made by all previous models.

In section~\ref{sec:theory} we lay out the theoretical background,
framework and the details of our novel quantum noise model. We 
illustrate in section~\ref{sec:influence}
 the effects of interferometer asymmetries using two 
examples of vastly different arm lengths, from the metre-scale
 Glasgow Sagnac speed meter proof-of-concept
  experiment to the potential speed meter implementation for the 
10km long Einstein Telescope on the other hand. We conclude with a summary
and outlook in section~\ref{sec:summary}.

\section{Analytical analysis of quantum noise in an imperfect and 
         asymmetric Sagnac speed meter}
\label{sec:theory}
\begin{figure}[h]
\begin{center}
	\includegraphics[width=\textwidth]{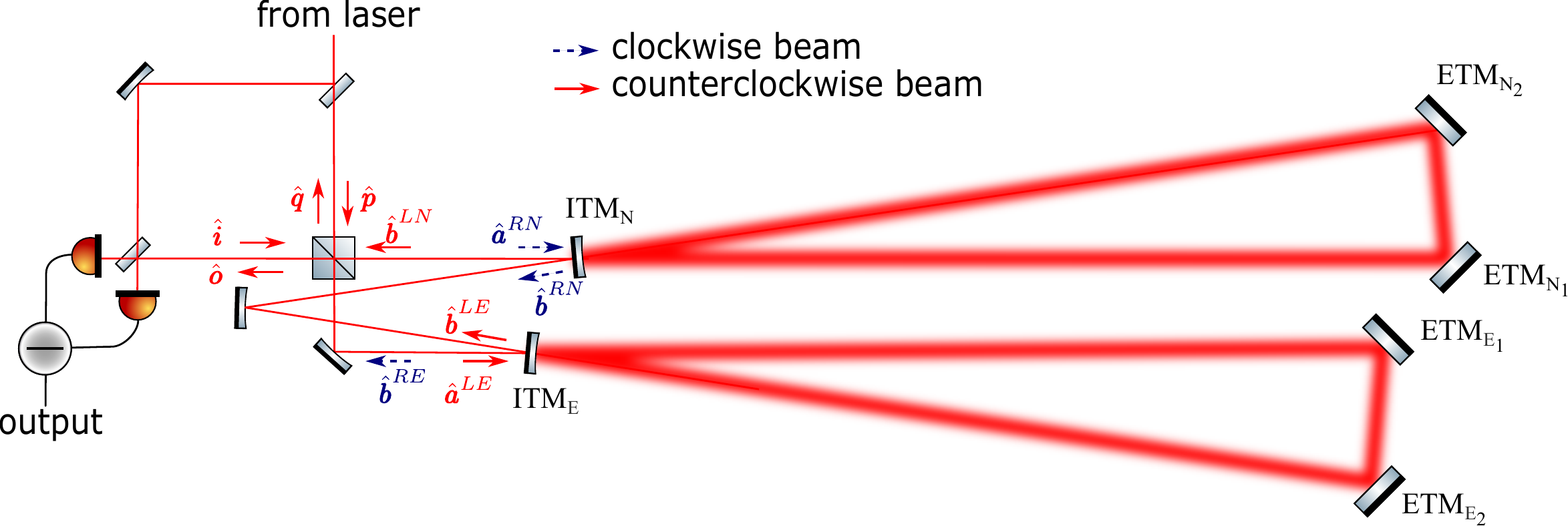}
\caption{\SD{Simplified optical layout of a proof-of-concept speed meter interferometer with ring arm cavities that is being built in the University of Glasgow and that we will base our treatment of quantum noise on.} Counterclockwise and clockwise beams as well as corresponding vacuum fields are 
denoted by blue (dashed) and red  arrows, respectively.}
\label{fig:Sag_layout}
\end{center}
\end{figure}

In this section, we calculate quantum noise limited sensitivity (or more accurately its spectral density) for an imbalanced Sagnac interferometer featuring arm cavities, as shown in Fig.~\ref{fig:Sag_layout}. \SD{This layout is chosen for a reason that it replicates the main design features of a Proof-of-Concept speed meter interferometer under construction at the University of Glasgow \cite{Graef14}. The most profound deviation of this setup from a large scale GW interferometer is that it has parallel arms, while the latter has orthogonal ones. However, we keep denoting the arms and all pertaining elements with the same letters $N$ and $E$ (meaning "north"- and "east"-bound arms, respectively) for compatibility with the earlier works \cite{Chen2002,Danilishin04,Liv.Rv.Rel.15.2012}. }
 
The main purpose of this section is to show what impact different imperfections have on the realistic Sagnac speed meter's ability to suppress quantum back-action noise if compared 
to Michelson interferometers. In particular, we study how the deviation of the beam splitter ratio from the ideal 50\%/50\% changes quantum noise. As well, the effect of non-identical arm cavities is 
considered. We study also the effect of optical loss in the elements of the core optics. 

\SD{Consider first the underlying principle that makes Sagnac interferometer a speed meter.  Indeed, visiting consequently both arms (see blue (dashed) and red arrows in Fig.~\ref{fig:Sag_layout}), two counter propagating light beams are reflected sequentially from both arm cavities thereby acquiring phase shifts proportional to the sum of arms length variations $\Delta x_{N,E}(t) \equiv (x_{\rm ETM}^{N,E}(t) - x_{\rm ITM}^{N,E}(t))$ (hereinafter I(E)TM stands for Input (End) Test Mass) of each of the cavities taken with time delay equal to average single cavity storage time $\tau_{\rm arm}$:
\begin{eqnarray}
    \delta\phi_R &\propto & \Delta x_N(t)+\Delta x_E(t+\tau_{\rm arm})\,,\\
    \delta\phi_L &\propto & \Delta x_E(t)+\Delta x_N(t+\tau_{\rm arm})\,.
\end{eqnarray}
After recombining at the beamsplitter and photo detection the output signal turns out to be proportional to the phase difference of the clockwise (R) and the counter clockwise (L) propagating light beams:
\begin{multline}\label{phi_speedmeter}
 \delta\phi_R - \delta\phi_L \propto [\Delta x_N(t)-\Delta x_N(t+\tau_{\rm arm})]-
   [\Delta x_E(t)-\Delta x_E(t+\tau_{\rm arm})]\propto\\
  \propto \Delta\dot x_N(t) - \Delta\dot x_E(t) + O(\tau_{\rm arm})
\end{multline}
that, for frequencies $\ll\tau_{\rm arm}^{-1}$, is proportional to relative rate of the interferometer arms length variation, \textit{i.e.} their relative speed. 

Note also that the optical paths of the two beams are absolutely identical irrespective of the difference in length of the two arms, if looked at on a time scale longer than $\tau_{\rm arm}$. Therefore, a Sagnac interferometer naturally keeps its output port dark at DC frequencies. It is only the dynamical change of the arms lengths faster than $\tau_{\rm arm}$ that leads to a non-zero signal at the output photodetector.
}

 We start the analysis of the scheme with choosing the proper notations for the optical fields on key elements of the interferometer. Unlike in Michelson interferometer, in Sagnac interferometers all photons 
pass through both arm cavities before recombining with a counter-propagating beam at the beam splitter. At the same time, the two light beams hit the cavity simultaneously,
 one coming directly from the beam splitter and the other one, that has just left the other arm. In notations of Chen's paper \cite{Chen2002}, quadrature operators of light entering 
 and leaving the arm can be marked with two indices $IJ$, \textit{e.g.} $a_c^{IJ}$, 
where $I$ stands for the either of two beams, L or R, and $J$ stands for the either of two arms ($J=E,N$). Here $R$ marks the light beam that first enters North arm and then travels the interferometer in 
the right direction (clockwise), and $L$ marks the beam travelling the interferometer in the opposite (counterclockwise) direction after entering the interferometer through the East arm. 

\subsection{Two-photon formalism for quantised light}

Quantum noise in interferometers originates from the quantum nature of light \cite{2003_PhysRevD.67.082001_Brag_et_al}. We will use the so called \textit{two-photon formalism} of Caves and Schumaker \cite{85a1CaSch,85a2CaSch} to describe 
quantised light and its quantum fluctuations in the most convenient manner for optomechanical displacement sensors, of which GW interferometers, including the Sagnac speed meter, make an important 
class.

The monochromatic electromagnetic wave with a  central frequency $\omega_0 = 2\pi c/\lambda_0$ and $\lambda_0$ its wavelength, can be characterised by its electric field strain. At an arbitrary point of 
space, characterised by the coordinate vector $\pmb{r} = \{x,\,y,\,z\}$, its space-time dependence can be written as:
  \begin{equation}
  \hat E(\pmb{r},\,t) = \mathcal{E}_0u(\pmb{r})\bigl[(A_c+\hat a_c(t))\cos\omega_0t + (A_s+\hat a_s(t))\sin\omega_0t\bigr]\,,
  \end{equation}
  where $\mathcal{E}_0 = \sqrt{4\pi\hbar\omega_0/(\mathcal{A} c)}$ with $\mathcal{A}$ the cross-section area of the light beam. Factor $u(\pmb{r})$ describes the spatial structure of the light field that may 
be quite peculiar. For our analysis, this factor is irrelevant as it does not influence quantum noise spectrum.  
Here we separated \textit{sine} and \textit{cosine} quadrature amplitudes in a classical (denoted by capital letters, $A_{c,s}$) and quantum fluctuation (small capped letters, $\hat a_{c,s}(t)$) parts, to track 
their propagation through the interferometer separately. Hence, the dynamics of the light field in the interferometer is reduced to the transformation of the 2-dimensional quadrature vectors:
\begin{equation}
	\pmb{A} = 
	\begin{bmatrix}
		A_c \\
		A_s
	\end{bmatrix}\,,\quad\mbox{and}\quad
	\hat{\pmb{a}} = 
	\begin{bmatrix}
		\hat a_c \\
		\hat a_s
	\end{bmatrix}\,.
\end{equation}

Usually, the analysis of light in linear optical devices is performed in Fourier domain. For the noise quadrature amplitudes it is done straightforwardly using the Fourier transform:
\begin{equation}
\hat a_{c,s}(t) = \int\limits_{-\infty}^\infty \frac{d\Omega}{2\pi}\,\hat a_{c,s}(\Omega)e^{-i\Omega t}\,,
\end{equation}
where $\Omega = \omega-\omega_0$ stands for the offset from the carrier frequency $\omega_0$. In the following  we will use only the Fourier picture and omit the argument $\Omega$ for convenience 
and clearer presentation.

\subsection{Input-Output relations for a linear optomechanical device.}

An optomechanical device can be characterised by a transformation that the mechanical motion of its parts imprints on the light passing through, or reflected from it. A Sagnac interferometer is a clear 
example of the optomechanical sensor. To calculate its quantum noise  we need to find how the input fluctuations of the light, characterised by quadrature amplitudes $\vq{a}^{in} = \{\hat{a}^{in}_c,\,\hat{a}
^{in}_s\}^{\rm T}$, get transformed by the interferometer into the output quadratures, $\vq{b}^{out} = \{\hat{b}^{out}_c,\,\hat{b}^{out}_s\}^{\rm T}$.  This task can be conveniently solved using a transfer 
matrix, or Input-Output (I/O) relations approach in the Fourier domain that can be written in general form as:
\begin{equation}\label{eq:IOlossless}
  \vq{b}^{out} = \tq{T}\cdot\vq{a}^{in} + \vs{R}_xx/x_{\rm SQL}\,,
\end{equation}
where
\begin{equation}
  \tq{T} \equiv
  \begin{bmatrix}
    T_{cc}(\Omega) & T_{cs}(\Omega)\\
    T_{sc}(\Omega) & T_{ss}(\Omega)
  \end{bmatrix}
\end{equation}
is the optical transfer matrix of the interferometer,
\begin{equation}
  \vs{R}_x \equiv
  \begin{bmatrix}
    R_{x,c}(\Omega)\\
    R_{x,s}(\Omega)
  \end{bmatrix}
\end{equation}
is an optical response of the interferometer on a mirror displacement with spectrum $x(\Omega)$, and
\begin{equation}\label{eq:xSQL}
  x_{\rm SQL} = \sqrt{\frac{2\hbar}{M\Omega^2}}
\end{equation}
is the free-mass amplitude spectral density of the Standard Quantum Limit in terms of mirror displacement for an interferometer with the
effective mechanical displacement mode mass $M$.

The output signal of the interferometer is usually contained in a photocurrent of a photodetector, or, if a more advanced readout technique is used, the difference current of a balanced homodyne detector, $\hat{i}^{out}_
\zeta$ that is proportional to the output light quadrature with the homodyne angle $\zeta$:
\begin{equation}
  \hat{i}^{out}_\zeta \propto \hat{b}^{out}_c\cos\zeta + \hat{b}^{out}_s\sin\zeta \equiv \vs{H}_\zeta^{\rm T}\cdot\vq{b}\,,\
  \vs{H}_\zeta\equiv
  \begin{bmatrix}
    \cos\zeta\\
    \sin\zeta
  \end{bmatrix}\,.
\end{equation}
The corresponding quantum noise spectral density in the desired units, \textit{e.g.} in units of displacement, can be obtained from the above using the following simple rule:
\begin{equation}\label{eq:SpDens_x}
  S^{x_-}(\Omega) = x^2_{\rm SQL}\frac{\vs[T]{H}_\zeta\cdot\tq{T}\cdot\tq{S}^{in}_{a}\cdot\tq[\dag]{T}\cdot\vs{H}_\zeta}{|\vs[T]{H}_\zeta\cdot\vs{R}_x|^2}
\end{equation}
where $\tq{S}^{in}_a$ is the spectral density matrix of the incident light, whose components are defined as:
\begin{equation}\label{eq:SpDens_a}
  2\pi\delta(\Omega-\Omega')\tq{S}^{in}_{a, ij}(\Omega) \equiv\\ \frac12\bra{in}\hat{a}^{in}_i(\Omega)(\hat{a}^{in}_j(\Omega'))^\dag+(\hat{a}^{in}_j(\Omega'))^\dag\hat{a}^{in}_i(\Omega)\ket{in}\,,
\end{equation}
where $\ket{in}$ is the quantum state of the light injected into the dark port of the interferometer and $(i,j) = (c,s)$  (see Sec. 3.3 in \cite{Liv.Rv.Rel.15.2012} for more details).

\subsection{Quantum noise in a real lossy interferometer}

The procedure described above is idealised because it neither takes into account optical losses and the associated additional quantum noise, 
nor the asymmetry present in any real balanced scheme.  In order to take those factors into account it is necessary to (i) consider arms of the interferometer separately and (ii) take into account optical 
loss in all elements of the scheme and add the corresponding incoherent noise terms into the inputs of the interferometer input-output (I/O) relations.

This leads to an expansion of the number of  inputs of the interferometer for, \textit{e.g.}, in a lossy system for each particular loss point one has to introduce a corresponding vacuum noise field according 
to the fluctuation-dissipation theorem \cite{PhysRev.83.34}. So, if one has a system with $N$ input fields, $\vq{a}^{in}_j$, and $M$ loss-associated noise fields, $\vq{n}_k$, the corresponding expression 
for the quantum noise spectral density will be just a trivial sum of spectral densities of the individual noise sources:
 \begin{equation}\label{eq:SpDens_x_loss}
  S^x(\Omega) = x^2_{\rm SQL}\frac{\sum\limits_{j=1}^N\vs[T]{H}_\zeta\cdot\tq{T}_j\cdot\tq{S}^{in}_{a_j}\cdot\tq[\dag]{T}_j\cdot\vs{H}_\zeta + \sum\limits_{k=1}^M\vs[T]{H}_\zeta\cdot\tq{N}_k\cdot\tq[\dag]
{N}_k\cdot\vs{H}_\zeta}{|\vs[T]{H}_\zeta\cdot\vs{R}_x|^2}
\end{equation}
 where $\tq{S}^{in}_{a_j}$ are (single-sided) spectral density matrices for all independent inputs, and we accounted for the special shape of a vacuum state spectral density matrix of the loss-associated 
vacuum fields, $\tq{S}^{in}_{n_k} = \tq{I}$ ---identity matrix (see, \textit{e.g.} Sec. 3.2.1 of \cite{Liv.Rv.Rel.15.2012}). 
 
\subsection{Input-output relations for a symmetric lossless Sagnac interferometer}
Before doing a full analysis of a lossy imperfect Sagnac interferometer, let us recall briefly the derivation of I/O-relations for a lossless Sagnac interferometer as is done in Chen's paper \cite{Chen2002} 
and keeping to his notations as described above:
\begin{eqnarray}
  b_c^{IJ} &=&   e^{2i\beta_{\rm arm}(\Omega)} a_c^{IJ}\,,\label{eq:I/O_arm_lossless_c} \\ 
  b_s^{IJ} &=&   e^{2i\beta_{\rm arm}(\Omega)} [a_s^{IJ} - \mathcal{K}_{\rm arm}^{IJ} a_c^{IJ}-\mathcal{K}_{\rm arm}^{\bar IJ} a_c^{\bar IJ}] + \\\nonumber
  & & e^{i\beta_{\rm arm}(\Omega)} \sqrt{2\mathcal{K}_{\rm arm}^{IJ}} \frac{\sqrt2x_J}{x_{\rm SQL}} \label{eq:I/O_arm_lossless_s}\label{eq:FP_arm_I/O-rel}
\end{eqnarray}
with $\bar I$ indicating the beam propagating in opposite direction with respect to $I$, \textit{i.e.} $\bar R = L$ and $\bar L = R$, and $x_J = x_J^{\rm ETM}-x_J^{\rm ITM}$ is the signal-induced arm 
elongation\footnote{Note that the factor $\sqrt2$ in front of the arm mechanical mode coordinate $x_J$ in Eq.~\eqref{eq:FP_arm_I/O-rel} is due to the difference between the effective mass of the arm, $\mu_{\rm arm}$, and that of the whole interferometer $M = \mu_{\rm arm}/2$, that enters the expression for $x_{\rm SQL}$ in Eq.~\eqref{eq:xSQL}.}. Here we introduce the optomechanical coupling coefficients, $\mathcal{K}^{IJ}_{\rm arm}$, for each beam separately. This notation helps us later on to account for asymmetries in the interferometer. For the definition of $\mathcal{K}^{IJ}_{\rm arm}$ we follow the Kimble \textit{et al.} paper \cite{02a1KiLeMaThVy}:
\begin{align}
\mathcal{K}_{\rm arm}^{IJ} &=  \frac{2\Theta^{IJ} \gamma_{\rm arm}}{\Omega^2(\gamma_{\rm arm}^2+\Omega^2)}\,,\quad\mbox{with}\quad \Theta^{IJ} = \dfrac{4\omega_0 P_c^{IJ}}{\mu_{\rm arm} cL}\,,
\label{eq:Karm}\\
 \beta_{\rm arm} &=  \mathrm{arctan}\dfrac{\Omega}{\gamma_{\rm arm}}\label{eq:Betaarm}\,,
\end{align}
where $\gamma_{\rm arm} = cT_{\rm ITM}/(4L)$ is the arm cavity half-banwidth, $P_c^{IJ}$ stands for optical power circulating in the arm in one direction, \textit{i.e.} in the R-beam, or in the L-beam, and 
$\mu_{\rm arm} = 2 M_{\rm ITM}M_{\rm ETM}/(M_{\rm ITM}+2 M_{\rm ETM})$ is the effective mass of the arm.

Now it is straightforward to derive full I/O-relations for a lossless symmetric Sagnac interferometer. In this case, the  optomechanical coupling coefficients are the same for all beams, \textit{i.e.} $
\mathcal{K}_{\rm arm}^{IJ} \equiv \mathcal{K}_{\rm arm}$. Then, using junction equations for the fields at the beam splitter:
\begin{align}\label{eq:I/O-rels_BS_lossless1}
  \vq{a}^{\rm RN} &= \dfrac{\vq{p}+\vq{i}}{\sqrt{2}}\,, & \vq{a}^{\rm LE} &= \dfrac{\vq{p}-\vq{i}}{\sqrt{2}}\,, & \vq{o} &= \dfrac{\vq{b}^{\rm LN}-\vq{b}^{\rm RE}}{\sqrt{2}}\,,
\end{align}
as well as continuity relations between the beams that leave one arm and enter the other:
\begin{align}\label{eq:I/O-rels_BS_lossless2}
\vq{a}^{\rm RE} &= \vq{b}^{\rm RN}\,, & \vq{a}^{\rm LN} &= \vq{b}^{\rm LE}\,.
\end{align}
one obtains:
\begin{equation}\label{eq:IOSaglossless}
  \begin{bmatrix} \hat{o}_c\\\hat{o}_s\end{bmatrix} =
  e^{2i\beta_{\rm sag}}\begin{bmatrix} 1 & 0\\-\mathcal{K}_{\rm sag} & 1\end{bmatrix}
  \begin{bmatrix} \hat{i}_c\\\hat{i}_s\end{bmatrix}+
  \begin{bmatrix}0\\\sqrt{2\mathcal{K}_{\rm sag}}\end{bmatrix}e^{i\beta_{\rm sag}}
\frac{x_{-}}{x_{\rm SQL}}\,,
\end{equation}
with the coupling constant $\mathcal{K}_{\rm sag}$ defined as:
\begin{equation}
  \mathcal{K}_{\rm sag} = 8 \mathcal{K}_{\rm arm} \sin^2\beta_{\rm arm} = \frac{4 \Theta \gamma_{\rm arm}}{(\Omega^2+\gamma_{\rm arm}^2)^2},
\end{equation}
and phase shift:
\begin{equation}\label{eq:SIphase}
  \beta_{\rm sag} = 2\beta_{\rm arm} + \frac{\pi}{2} \,.
\end{equation}
Here we define the differential mechanical mode of the  interferometer as $x_{-} = x_N-x_E$ (the common mode is defined by analogy as $x_{+} = x_N+x_E$).

The noise transfer matrix and signal response vector for this case have a particularly concise form:
\begin{equation}\label{eq:SI_T_and_t}
  \tq{T} = -e^{2i\beta_{\rm sag}}\begin{bmatrix} 1 & 0\\-\mathcal{K}_{\rm sag} & 1\end{bmatrix}\,,\quad\vs{R} = e^{i\beta_{\rm sag}} \begin{bmatrix}0\\\sqrt{2\mathcal{K}_{\rm sag}}\end{bmatrix}\,.
\end{equation}
Therefore one gets the following simple expression for the spectral density of the quantum noise limited sensitivity
of the zero-area Sagnac interferometer  (it is the same for all tuned interferometers with a balanced homodyne readout of quadrature $b_\zeta$ and a vacuum state at the dark port, save to the expression 
for $\mathcal{K}$):
\begin{equation}\label{eq:Sx_plain}
   S^{x_-} = \frac{x^2_{\rm SQL}}{2}\left\{\frac{\left[\mathcal{K}_{\rm sag}-\cot\zeta\right]^2+1}{\mathcal{K}_{\rm sag}}\right\}\,.
\end{equation}  

\subsection{Asymmetric beam splitter}
The main asymmetry one can think of in a Sagnac interferometer is the non-perfect splitting ratio of the main beam splitter (BS) resulting in an imbalance of the power in the two light beams propagating in 
opposite directions. As our analysis demonstrates below, this imbalance leads to a dramatic increase of the residual radiation pressure noise, amounting to a steeper rise of the quantum noise towards 
lower frequencies, $S^{\rm r.p.}_x\propto f^{-6}$, than that of a Michelson interferometer. 
 
In order to account for this asymmetry in our quantum noise calculations let us define the BS symmetry offset, $\eta_{\rm BS}$, through the BS power reflectivity, $R_{\rm BS}$, and transmissivity, $T_{\rm 
BS}$, as:
\begin{align}
\sqrt{R_{\rm BS}} &= \dfrac{1+\eta_{\rm BS}}{\sqrt{2}}\,, & \sqrt{T_{\rm BS}} &= \dfrac{1-\eta_{\rm BS}}{\sqrt{2}}\,.
\end{align}
Then the Sagnac I/O-relations with an asymmetric BS read (see Fig.~\ref{fig:BS} for field operator notations):
\begin{align*}
&\vq{o} = \dfrac{\vq{b}^{LN}-\vq{b}^{RE}}{\sqrt2} + \eta_{\rm BS}\dfrac{\vq{b}^{LN}+\vq{b}^{RE}}{\sqrt2}\,,
&\vq{q} = \dfrac{\vq{b}^{LN}+\vq{b}^{RE}}{\sqrt2} - \eta_{\rm BS}\dfrac{\vq{b}^{LN}-\vq{b}^{RE}}{\sqrt2}\,,\\
&\vq{a}^{RN} = \dfrac{\vq{p}+\vq{i}}{\sqrt2} + \eta_{\rm BS}\dfrac{\vq{p}-\vq{i}}{\sqrt2}\,, 
&\vq{a}^{LE} = \dfrac{\vq{p}-\vq{i}}{\sqrt2} - \eta_{\rm BS}\dfrac{\vq{p}+\vq{i}}{\sqrt2}\,,\\
\end{align*}

Using these expressions one can immediately see that the classical amplitudes of the two beams, leaving the beam-splitter, are uneven, \textit{i.e.} $A^{RN} = P(1+\eta_{\rm BS})/\sqrt{2}$ and $A^{LE} = 
P(1-\eta_{\rm BS})/\sqrt{2}$ ($P$ is a classical amplitude of pump field , $\vq{p}$, and we assume no classical component for the field entering through the dark port, $I=0$). Therefore, the same is true for the intracavity fields and thereby for the optomechanical 
coupling factors $\mathcal{K}_{\rm arm}^{IJ}$, which can now be written as:
\begin{align}\label{eq:asym_BS_offset}
\mathcal{K}_{\rm arm}^{RN} &=  \mathcal{K}_{\rm arm}^{RE} = \mathcal{K}_{\rm arm}(1+\eta_{\rm BS})^2\,, & \mathcal{K}_{\rm arm}^{LE} &=  \mathcal{K}_{\rm arm}^{LN} = \mathcal{K}_{\rm arm}(1-
\eta_{\rm BS})^2\,,
\end{align}
which indicates the imbalance in the radiation pressure force responsible for the effect we are describing in this subsection.

The I/O-relations for the Sagnac interferometer with an asymmetric beam splitter can be written as:
\begin{equation}\label{eq:IOasymBS}
  \vq{o} = \tq{T}_i^{\rm asym.\,BS}\vq{i} + \tq{T}_p^{\rm asym.\,BS}\vq{p}+ \vs{R}_{-}x_{-}/x_{\rm SQL} + \vs{R}_{\rm +}x_{+}/x_{\rm SQL}\,.
\end{equation}
where the quantum noise transfer matrices read:
\begin{align*}
\tq{T}_i^{\rm asym.\,BS} &=
(1-\eta^2_{\rm BS})e^{2i\beta_{\rm sag}}
\begin{bmatrix}
1 & 0\\
-\bigl[\mathcal{K}^{\rm sag}_{\rm sym} + \eta_{\rm BS}^2\mathcal{K}^{\rm sag}_{\rm asym}\bigr] & 1
\end{bmatrix}\,, \\
\tq{T}_p^{\rm asym.\,BS} &=
2\eta_{\rm BS}e^{2i\beta_{\rm sag}}
\begin{bmatrix}
1 & 0\\
-\frac12\bigl[(1+3\eta_{\rm BS}^2)\mathcal{K}^{\rm sag}_{\rm sym} + (3+\eta_{\rm BS}^2)\mathcal{K}^{\rm sag}_{\rm asym}\bigr] & 1
\end{bmatrix}
\end{align*}
and where we define the new phase shift, $\beta_{\rm sag}$, and the symmetric and asymmetric components of the optomechanical coupling as:
\begin{align}
\mathcal{K}^{\rm sag}_{\rm sym} &=  4\mathcal{K}_{\rm arm}\sin^2\beta_{\rm arm} = \dfrac{8\Theta\gamma_{\rm arm}}{(\gamma_{\rm arm}^2+\Omega^2)^2}\,, \\ \mathcal{K}^{\rm sag}_{\rm asym} &=  
4\mathcal{K}_{\rm arm}\cos^2\beta_{\rm arm} = \dfrac{8\Theta\gamma^3_{\rm arm}}{\Omega^2(\gamma_{\rm arm}^2+\Omega^2)^2}\,,
\end{align}
and $\beta_{\rm sag} = 2\beta_{\rm arm}+\pi/2$.

Quite expectedly, an asymmetry of the beam splitter results in the common mode ($x_{+}$), signal showing up at the output port on a par with the differential mode. The two response functions for the 
cARM and the dARM signal read:
\begin{align}
\vs{R}_{\rm -} &=  e^{i\beta_{\rm sag}}(1+\eta_{\rm BS}^2)\sqrt{2\mathcal{K}^{\rm sag}_{\rm sym}}
\begin{bmatrix}
0\\
1
\end{bmatrix}\,, 
& \vs{R}_{\rm +} &=  2\eta_{\rm BS}e^{2i\beta_{\rm arm}}\sqrt{2\mathcal{K}^{\rm sag}_{\rm asym}}
\begin{bmatrix}
0\\
1
\end{bmatrix}\,.
\end{align}

It is now straightforward to the calculate spectral density of quantum noise in units of dARM displacement, using Eq.~\eqref{eq:SpDens_x}:
\begin{multline}
S^{x_-}_{\rm asym. BS} = \dfrac{x_{\rm SQL}^2}{2\mathcal{K}^{\rm sag}_{\rm sym}}\Biggl\{ \left(\dfrac{1-\eta_{\rm BS}^2}{1+\eta_{\rm BS}^2}\right)^2\left(1+\bigl[\mathcal{K}^{\rm sag}_{\rm sym} + \eta_{\rm BS}^2\mathcal{K}^{\rm sag}_{\rm asym}-\cot\zeta\bigr]^2\right) +\\
\left(\dfrac{2\eta_{\rm BS}}{1+\eta_{\rm BS}^2}\right)^2\left(1+\left[\frac12\bigl[(1+3\eta_{\rm BS}^2)\mathcal{K}^{\rm sag}_{\rm sym} + (3+\eta_{\rm BS}^2)\mathcal{K}^{\rm sag}_{\rm asym}\bigr]-\cot\zeta\right]^2 \right)
\Biggr\}
\end{multline}
Despite relative complexity of this formula, the origin of predicted steep rise of the quantum noise at low frequencies can be easily seen through. It directly follows from behaviour of  $\mathcal{K}^{\rm sag}_{\rm asym}$ and $\mathcal{K}^{\rm sag}_{\rm sym}$ at low frequencies $\Omega\ll\gamma_{\rm arm}$. Since $\mathcal{K}^{\rm sag}_{\rm sym}(\Omega\to0)\propto const$, $\mathcal{K}^{\rm sag}_{\rm asym}(\Omega\to0)\propto \Omega^{-2}$ and $x_{\rm SQL}^2\propto\Omega^{-2}$, the terms responsible for $\propto\Omega^{-6}$ rise are those proportional to $(\mathcal{K}^{\rm sag}_{\rm ssym})^2\propto \Omega^{-4}$ inside the braces. Together with $x^2_{\rm SQL}\propto\Omega^{-2}$ it gives the predicted behaviour. 


\subsection{Losses in the arm cavities}

The next important source of imperfection in a Sagnac interferometer is optical loss in the arm cavities. 

\begin{figure}[h]
\begin{center}
	\includegraphics[width=.7\textwidth]{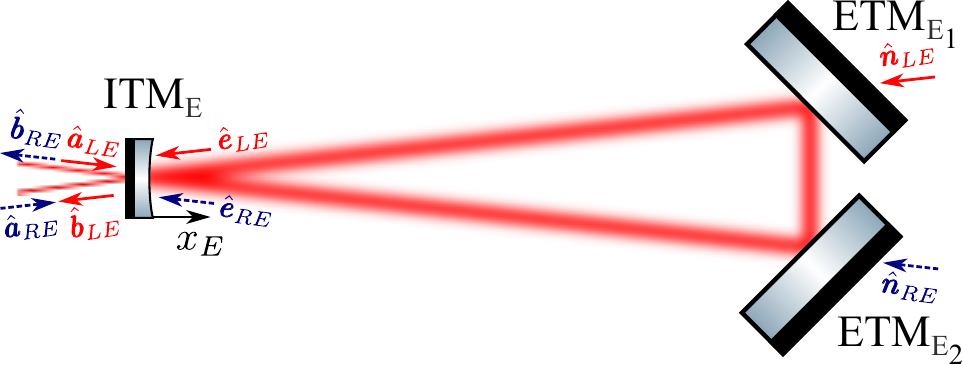}
\caption{Schematic of a Sagnac ring arm cavity with marked input and output fields. The ``east'' arm cavity is chosen for definiteness.}
\label{fig:arm_cav}
\end{center}
\end{figure}

Each arm cavity of the Sagnac interferometer can be considered as a Fabry-P\'erot--type ring cavity with movable mirrors as shown in Fig.~\ref{fig:arm_cav}. To account for losses in the arms we have to 
introduce additional vacuum fields in accordance with the fluctuation-dissipation theorem \cite{PhysRev.83.34}. For all practical purposes it is sufficient to model it by attributing an additional transmissivity 
to the end mirrors (ETMs), $T_{\rm loss}$. In this case, the general structure of the I/O-relations will remain similar to  Eqs.~\eqref{eq:I/O_arm_lossless_c} and \eqref{eq:I/O_arm_lossless_s}, but with 
additional vacuum noise fields originating from loss:
%
\begin{multline}\label{eq:I/O_arm_lossy}
\vq{b}^{IJ} = \tq{T}^{IJ}_{\rm arm}\cdot\vq{a}^{IJ} + \tq{N}^{IJ}_{\rm arm}\cdot\vq{n}^{IJ} + \tq{T}_{\rm arm,\ r.p.}^{\bar{I}J}\cdot\vq{a}^{\bar{I}J} +\tq{N}_{\rm arm,\ r.p.}^{\bar{I}J}\cdot\vq{n}_{\bar{I}J} + \vs{R}
^{IJ}_{\rm arm}\frac{x_J}{x_{\rm SQL}}\,,
\end{multline}
where $\vq{a}^{IJ}$ and $\tq{T}^{IJ}_{\rm arm}$ stand for for vacuum field entering the arm cavity through the ITM and its transfer matrix, $\tq{T}_{\rm arm,\ r.p.}^{\bar{I}J}$, represents a part of the full 
transfer matrix resulting from radiation pressure created by the counter propagating light beam, $\vq{n}^{IJ}$ and $\tq{N}^{IJ}_{\rm arm}$ stand for the loss-associated vacuum field entering the arm cavity 
through the ETM and its transfer matrix, $\tq{N}_{\rm arm,\ r.p.}^{\bar{I}J}$ is the radiation pressure component of the latter, while  $ \vs{R}^{IJ}$ is the cavity response to the mirror displacement. Entry 
points of all participating vacuum fields are shown schematically in Fig.~\ref{fig:arm_cav}.

Optical loss in the Sagnac interferometer manifests itself in two ways that conspire to undermine the radiation pressure suppression effect of the speed meter. Firstly, the power of the light beam when it 
leaves the first arm cavity towards the second cavity is reduced by a factor $\epsilon_{\rm arm} = T_{\rm loss}/(T_{\rm ITM}+T_{\rm loss})$, and therefore the radiation pressure force it creates in the 
second cavity is less than that in the first one. As a result, the perfect subtraction of radiation pressure forces becomes impossible. Secondly, the additional uncorrelated vacuum noise that accompanies 
the light beam at its second reflection of the arm cavity, right before the recombination at the beam splitter, creates an uncompensated radiation pressure force akin to that of a Michelson interferometer. 
These two effects together are responsible for the rise of the quantum noise at the low frequencies. 

In order to distinguish the symmetric loss effect from  the effect of imbalance, it is reasonable to represent the cavity mirror parameters as a sum of symmetric and anti-symmetric components in the 
following way:
\begin{align*}
T^J_{\rm ITM} &= T_{\rm ITM} \pm \delta T_{\rm ITM}/2\,, \quad\Leftrightarrow\quad &T_{\rm ITM}=\dfrac{T^N_{\rm ITM}+T^E_{\rm ITM}}{2}\,,\quad \delta T_{\rm ITM}=T^N_{\rm ITM}-T^E_{\rm ITM}\,.
\end{align*}
$T^J_{\rm loss}$ can be represented in a similar way. Then one can represent all the arm-related imperfections in terms of four parameters, namely:
\begin{enumerate}
\item
average bandwidth, $\gamma_{\rm arm} = \dfrac{c(T_{\rm ITM}+T_{\rm loss})}{4L}$;
\item
its imbalance $\delta\gamma = \dfrac{c(\delta T_{\rm ITM}+\delta T_{\rm loss})}{4L}$;
\item
average fractional loss of photons per round trip per cavity, $\epsilon_{\rm arm} = \dfrac{T_{\rm loss}}{T_{\rm ITM}+T_{\rm loss}}$;
\item
and associated imbalance $\delta\epsilon_{\rm arm} \simeq \dfrac{\delta T_{\rm loss}}{T_{\rm ITM}+T_{\rm loss}}$. 
\end{enumerate}
Another common feature of these imperfections, confirmed by numerical estimates based on general treatment outlined in \ref{app_sec:I/O_relations_imp_sag} is that their impact is noticeable only at 
frequencies well below the arm cavity bandwidth, \textit{i.e.} for $\Omega\ll\gamma_{\rm arm}$. Keeping this in mind and using the introduced parameters, one can rewrite optomechanical coupling factors 
for the arms, defined in \eqref{eq:Karm} as (we set $\eta_{\rm BS}=0$ here for simplicity and to isolate the effect of the arms from that of the BS):
\begin{subequations}
\begin{align}
\mathcal{K}^{RN}_{\rm arm} &= \mathcal{K}_{\rm arm} \left\{1-\frac{\delta\gamma}{\gamma_{\rm arm}}-\frac{\delta\epsilon_{\rm arm}}{2}-\epsilon_{\rm arm}\bigl(1-\frac{\delta\gamma}{\gamma_{\rm arm}}
\bigr)\right\}\,,\\
\mathcal{K}^{LE}_{\rm arm} &= \mathcal{K}_{\rm arm} \left\{1+\frac{\delta\gamma}{\gamma_{\rm arm}}+\frac{\delta\epsilon_{\rm arm}}{2}-\epsilon_{\rm arm}\bigl(1+\frac{\delta\gamma}{\gamma_{\rm arm}}
\bigr)\right\}\,,\\
\mathcal{K}^{RE}_{\rm arm} &= \mathcal{K}_{\rm arm} \left\{1-\frac{\delta\gamma}{\gamma_{\rm arm}}-\epsilon_{\rm arm}\bigl(2-\frac{\delta\gamma}{\gamma_{\rm arm}}+\frac{\delta\epsilon_{\rm arm}}
{2}\bigr)+\epsilon_{\rm arm}^2\right\}\,,\\
\mathcal{K}^{LN}_{\rm arm} &= \mathcal{K}_{\rm arm} \left\{1+\frac{\delta\gamma}{\gamma_{\rm arm}}-\epsilon_{\rm arm}\bigl(2+\frac{\delta\gamma}{\gamma_{\rm arm}}-\frac{\delta\epsilon_{\rm arm}}
{2}\bigr)+\epsilon_{\rm arm}^2\right\}\,.
\end{align}
\end{subequations}
One can see that the effect of symmetric loss on the optomechanical interaction ($\delta\gamma=\delta\epsilon_{\rm arm}=0$) is reduced to the multiplication of the  loss-free $\mathcal{K}_{\rm }$ by $(1-
\epsilon_{\rm arm})$ in the first passage of the beam through the arm cavity ($RN$ and $LE$ beams), and by $(1-\epsilon_{\rm arm})^2$ in the second passage ($RE$ and $LN$ beams), which is 
expectable. The phase shift $\beta_{\rm arm}$ is also modified by loss and asymmetry via $\gamma^J_{\rm arm}\to \gamma^J_{\rm arm}(1+\epsilon_{\rm arm}\pm\delta\epsilon_{\rm arm}/2)$, but the 
increment is a second order correction $\sim\mathcal{O}(\epsilon_{\rm arm}\Omega/\gamma_{\rm arm})$ and therefore omitted.

Inserting these expressions into Eqs.~\eqref{app_eq:I/O_arms_TMs_res} for the transfer matrices of lossy arms and then into \eqref{app_eq:I/O-rels_arms_res}, one gets the  I/O-relations  for lossy arms 
of the form shown in Eq.~\eqref{eq:I/O_arm_lossy}. Using symmetric beam splitter relations (refer to Eq.~\eqref{eq:I/O-rels_BS_lossless1} and Eq.~\eqref{eq:I/O-rels_BS_lossless2}), one can finally obtain 
the I/O-relations for a Sagnac interferometer with loss in the arms and get the expression for the spectral density, which is rather involved. However, the general structure of it can be represented as follows:
\begin{multline}\label{eq:Sh_sag_loss}
S^{x_-}_{\rm loss} = S^{x_-}+\frac{x^2_{\rm SQL}}{2}\mathcal{K}_{\rm arm}\Biggl\{L_{sym}(\epsilon_{\rm arm},\,\epsilon_{\rm arm}^2,\ldots) +\\
+ L_{asym}\left(\delta\epsilon_{\rm arm}^2,\,\frac{\delta\gamma\delta\epsilon_{\rm arm}}{\gamma_{\rm arm}}, \left(\frac{\delta\gamma}{\gamma_{\rm arm}}\right)^2\right) + \mathcal{O}(\delta\epsilon_{\rm 
arm}^3,(\delta\gamma/\gamma_{\rm arm})^3,\ldots)\Biggr\}\,,
\end{multline} 
where $S^{x_-}$ stands for the lossless Sagnac interferometer quantum noise spectral density of Eq.~\eqref{eq:Sx_plain} and both, $L_{sym}$ and $L_{asym}$, are linear functions. As one can see, the 
influence of loss in general is dictated by the factor $\mathcal{K}_{\rm arm}$ in front of the bracket which rises as $1\Omega^2$ at low frequencies and combined with $x_{\rm SQL}^2\propto 1/\Omega^2$ 
gives exactly the Michelson-like raise of quantum noise at low frequencies. 

\begin{figure}[h]
\begin{center}
	\includegraphics[width=.9\textwidth]{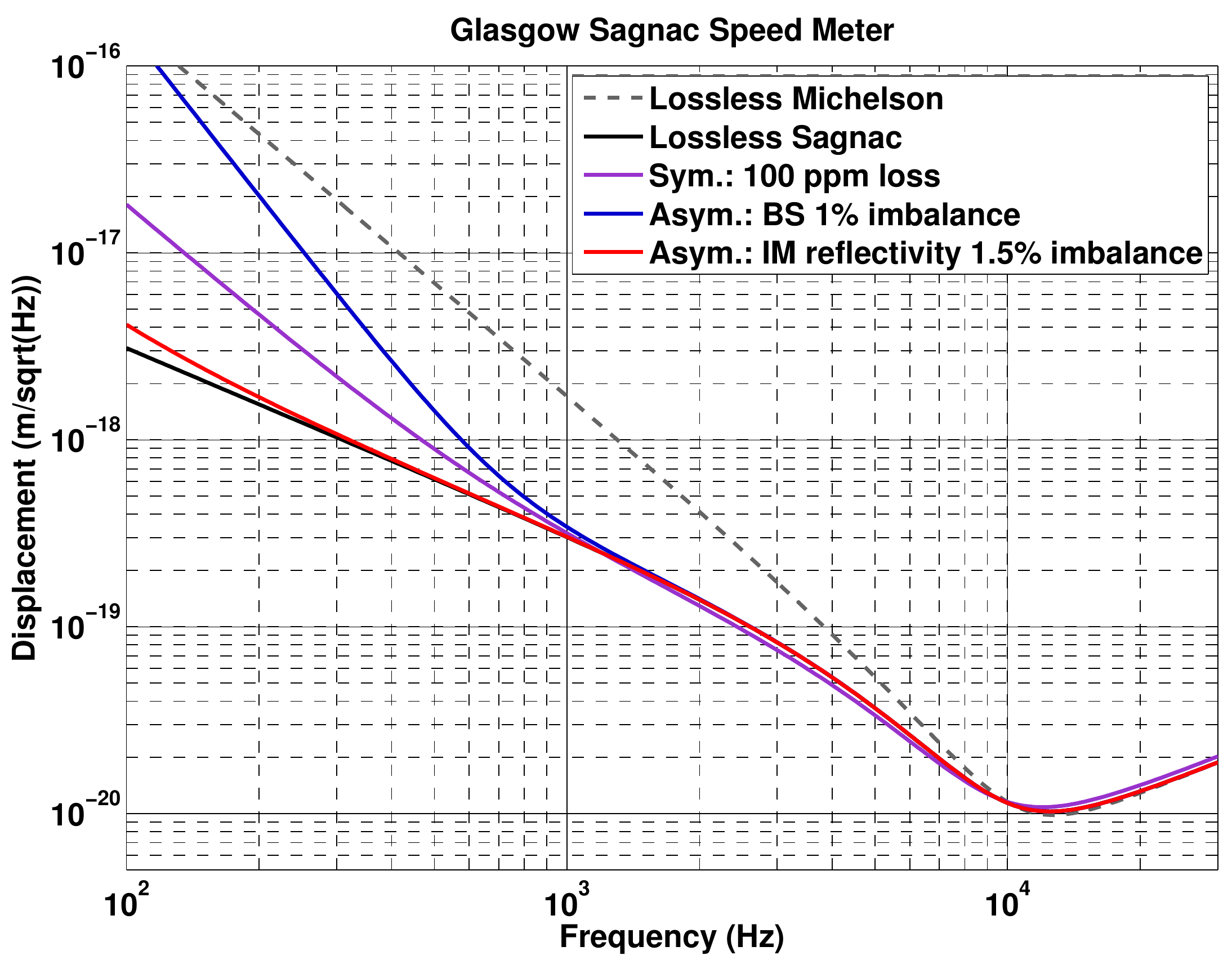}
\caption{Spectral density plots for a table-top Sagnac interferometer with parameters given in Table.~\ref{tab:etlf_params}. Here we demonstrate what impact different imperfections have on the quantum 
noise sensitivity of the interferometer. All plots are drawn for phase quadrature readout, \textit{i.e.} for homodyne angle $\zeta=\pi/2$. Ideal Michelson interferometer parameters match those of the corresponding ideal Sagnac interferometer, shown in the same plot.}
\label{fig:AS_Sag_SpDens}
\end{center}
\end{figure}

Asymmetries in the arms have a second-order influence, as indicated by the powers of the arguments of $L_{asym}$. In contrast symmetric loss has a first-order contribution to the total quantum noise of 
a Sagnac interferometer. These trends are demonstrated in Fig.~\ref{fig:AS_Sag_SpDens}, and the detailed behaviour of quantum noise as a function of symmetric loss, $\epsilon_{\rm arm}$, is shown 
in Fig.~\ref{fig:SSM_vs_ET_symmetric_loss}. The influence of asymmetry of the ITM transmissivities, or $\delta\gamma/\gamma_{\rm arm}$, 
is shown in Fig.~\ref{fig:SSM_vs_ET_ITM_transmissivity_imbalance}. The asymmetric loss, $\delta\epsilon_{\rm arm}$, has a similarly weak impact.

\subsection{General treatment of quantum noise of asymmetric Sagnac interferometer}

\begin{figure}[h]
\begin{center}
	\includegraphics[width=.35\textwidth]{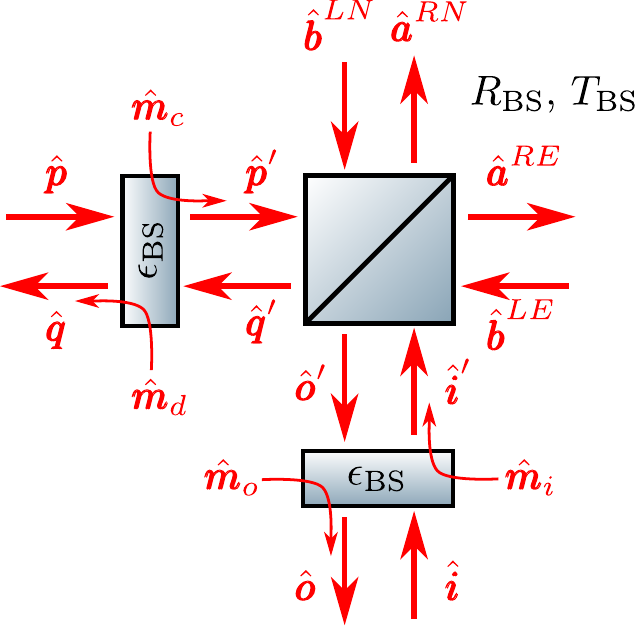}
\caption{Schematic of a lossy beamsplitter and its I/O-relations and fields.}
\label{fig:BS}
\end{center}
\end{figure}

For proper treatment of quantum noise in an asymmetric Sagnac, we need to specify the I/O-relations for a lossy beam splitter with arbitrary splitting ratio. The scheme of such a device with all the input 
and output fields is shown in Fig.~\ref{fig:BS}, and the relations between them read:
\begin{subequations}
\begin{align}
\vq{o}' &= -\sqrt{R_{\rm BS}} \vq{b}^{RE} + \sqrt{T_{\rm BS}} \vq{b}^{LN}\, , \label{eq:bs_o_lossless}\\
\vq{d}' &= \sqrt{T_{\rm BS}} \vq{b}^{RE} + \sqrt{R_{\rm BS}} \vq{b}^{LN}\, , \label{eq:bs_d_lossless}\\
\vq{a}^{RN} &= \sqrt{T_{\rm BS}} \vq{i}'  + \sqrt{R_{\rm BS}} \vq{c}' \, ,\label{eq:bs_aRN_lossless}\\ 
\vq{a}^{LE} &= -\sqrt{R_{\rm BS}} \vq{i}' + \sqrt{T_{\rm BS}} \vq{c}'\,. \label{eq:bs_aLE_lossless}
\end{align}
\end{subequations}

Optical loss can be included in the above I/O-relations following a standard procedure of complementing the lossless element with two virtual splitters of transmissivity $1-\epsilon_{\rm BS}$ and reflectivity $
\epsilon_{\rm BS}$, with the latter standing for average photon loss due to absorption in the beam splitter (see Fig.~\ref{fig:BS} for notations). This allows for additional incoherent vacuum fields associated with 
the loss to be included in the description as per fluctuation-dissipation theorem \cite{PhysRev.83.34}. As a result, we get the full I/O-relations for a lossy beam splitter in the following form:
\begin{subequations}\label{eq:BS_lossy}
\begin{align}
\vq{o} &= \sqrt{1-\epsilon_{\rm BS}} (-\sqrt{R_{\rm BS}} \vq{b}^{RE} + \sqrt{T_{\rm BS}} \vq{b}^{LN}) + \sqrt{\epsilon_{\rm BS}} \vq{m}_o\, , \label{eq:bs_o}\\
\vq{d} &= \sqrt{1-\epsilon_{\rm BS}} (\sqrt{T_{\rm BS}} \vq{b}^{RE} + \sqrt{R_{\rm BS}} \vq{b}^{LN}) + \sqrt{\epsilon_{\rm BS}} \vq{m}_d\, , \label{eq:bs_d}\\
\vq{a}^{RN} &= \sqrt{T_{\rm BS}} (\sqrt{1-\epsilon_{\rm BS}}\vq{i} + \sqrt{\epsilon_{\rm BS}} \vq{m}_i) + \sqrt{R_{\rm BS}} (\sqrt{1-\epsilon_{\rm BS}}\vq{c} + \sqrt{\epsilon_{\rm BS}} \vq{m}_c) \, ,\label{eq:bs_aRN}\\ 
\vq{a}^{LE} &= -\sqrt{R_{\rm BS}} (\sqrt{1-\epsilon_{\rm BS}}\vq{i} + \sqrt{\epsilon_{\rm BS}} \vq{m}_i) + \sqrt{T_{\rm BS}} (\sqrt{1-\epsilon_{\rm BS}}\vq{c} + \sqrt{\epsilon_{\rm BS}} \vq{m}_c)\,. \label{eq:bs_aLE}
\end{align}
\end{subequations}
One can check that substitutions $R'_{\rm BS}\to(1-\epsilon_{\rm BS})R_{\rm BS}$ and $T'_{\rm BS}\to(1-\epsilon_{\rm BS})T_{\rm BS}$  lead to a more traditional form of the I/O-relations where $R'_{\rm BS}+T'_{\rm BS}+\epsilon_{\rm BS} = 1$, 
while the meaning remains unchanged. 

Using these relations and the expressions for transfer matrices and response functions of a lossy arm cavity, derived in \ref{app_sec:I/O_relations_imp_sag}, we can calculate I/O-relations for a full Sagnac 
interferometer in the form:
\begin{align}
 \vq{o} &= \tq{T}^i_{\rm sag}\cdot\vq{i} + \tq{T}^p_{\rm sag}\cdot\vq{p} + \sum\limits_{\substack{I=L,R\\ J=N,E}}\tq{N}^{IJ}_{\rm sag}\cdot\vq{n}_{IJ} \sum\limits_{k=i,p}\tq{M}^{k}_{\rm sag}\cdot\vq{m}_{k} + 
\vb{R}^{+}_{\rm sag}x_++\vb{R}^{-}_{\rm sag}x_-\,.
\end{align}
Using this expression one can finally arrive at the general formula for quantum noise spectral density:
 \begin{multline}\label{eq:SpDens_x_asSag}
  S^x(\Omega) = \frac{x^2_{\rm SQL}}{|\vs[T]{H}_\zeta\cdot\vs{R}^-_{\rm sag}|^2}\Bigl\{\vs[T]{H}_\zeta\cdot[\tq{T}^i_{\rm sag}\cdot\tq{S}^{in}_{i}\cdot(\tq{T}^i_{\rm sag})^\dag+\tq{T}^p_{\rm sag}\cdot(\tq{T}
^p_{\rm sag})^\dag]\cdot\vs{H}_\zeta + 
 \\ + \sum\limits_{\substack{I=L,R\\ J=N,E}}\vs[T]{H}_\zeta\cdot\tq{N}^{IJ}_{\rm sag}\cdot(\tq{N}^{IJ}_{\rm sag})^\dag\cdot\vs{H}_\zeta+\sum\limits_{k=i,p}\vs[T]{H}_\zeta\cdot\tq{M}^{k}_{\rm sag}\cdot(\tq{M}
^{k}_{\rm sag})^\dag\cdot\vs{H}_\zeta\Bigr\}\,.
\end{multline}

\section{Influence on the performance of a small and a large scale speed meter}
\label{sec:influence}
In this section we present potential applications for
the model developed in the previous section. 
 We chose two specific Sagnac speed meter 
interferometer configurations as examples: the metre-scale experiment 
currently under construction at Glasgow and a large-scale 
configuration with parameters suitable for implementation as 
the low frequency interferometer \cite{Hild10} as part of the planned
 Einstein Telescope (ET) observatory. Both examples are 
 based on Sagnac interferometers employing  ring cavities 
 in the arms and a homodyne readout. Neither configuration discussed
 here contains recycling techniques or squeezed light injection. 
 
 \SD{
 Worth noting also is that all the plots presented  herein are drawn in assumption that we measure a phase quadrature of the outgoing light. 
 This is by no means an optimal regime for the speed meter in terms of surpassing the SQL (see \textit{e.g.} Sec. 6.2 of \cite{Liv.Rv.Rel.15.2012}), and much better sub-SQL
  sensitivity can be achieved with optimally tuned readout phase of Sagnac interferometer. The main goal of this paper is to demonstrate that even with imperfections, the Sagnac interferometer has significant advantage over the Michelson interferometer at low frequencies. To facilitate the reader in getting this message, we placed in all sensitivity plots in this article the sensitivity curves of an ideal (lossless and symmetric) Michelson interferometers with parameters equivalent to the corresponding ideal Sagnac interferometers as a yardstick.
  }

\begin{table}[htp]
\caption{Key parameters used to model the quantum-noise 
limited sensitivity of the Glasgow Sagnac speed meter proof of principle 
experiment and a large scale ET-LF like Sagnac configuration.}
\begin{tabular}{p{4.8cm}|p{3.5cm}|p{3cm}}
\hline
\textbf{Parameter} & \textbf{Glasgow speed meter} & \textbf{ET speed meter}\\
\hline\hline
Power incident on BS & 1.7\,W& $45.73$\,W \\
Laser wavelength & 1064\,nm & $1064$\,nm \\
Arm cavity round trip length & 2.83\,m & $2 \times 10^4$\,m \\
ITM mass & 0.85\,g & 211\,kg \\
ETM mass & 100\,g& 211\,kg \\
ITM transmissivity & 700\,ppm&  10000\,ppm \\
Photodiode efficiency & 95\,\% & 95\,\% \\
Beam splitter loss & $1000$\,ppm & $1000$\,ppm \\
 \hline
\end{tabular}

\label{tab:etlf_params}
\end{table}

The Glasgow Sagnac speed meter aims to demonstrate the 
back action reduction of a speed meter compared to a Michelson
interferometer with similar parameters. A detailed description 
of the experimental set up can be found in \cite{Graef14}. The 
most important parameters of this configuration are listed 
in the central column of Table~\ref{tab:etlf_params}. 

\begin{figure}[h]
  \centering
  \includegraphics[width=\textwidth]{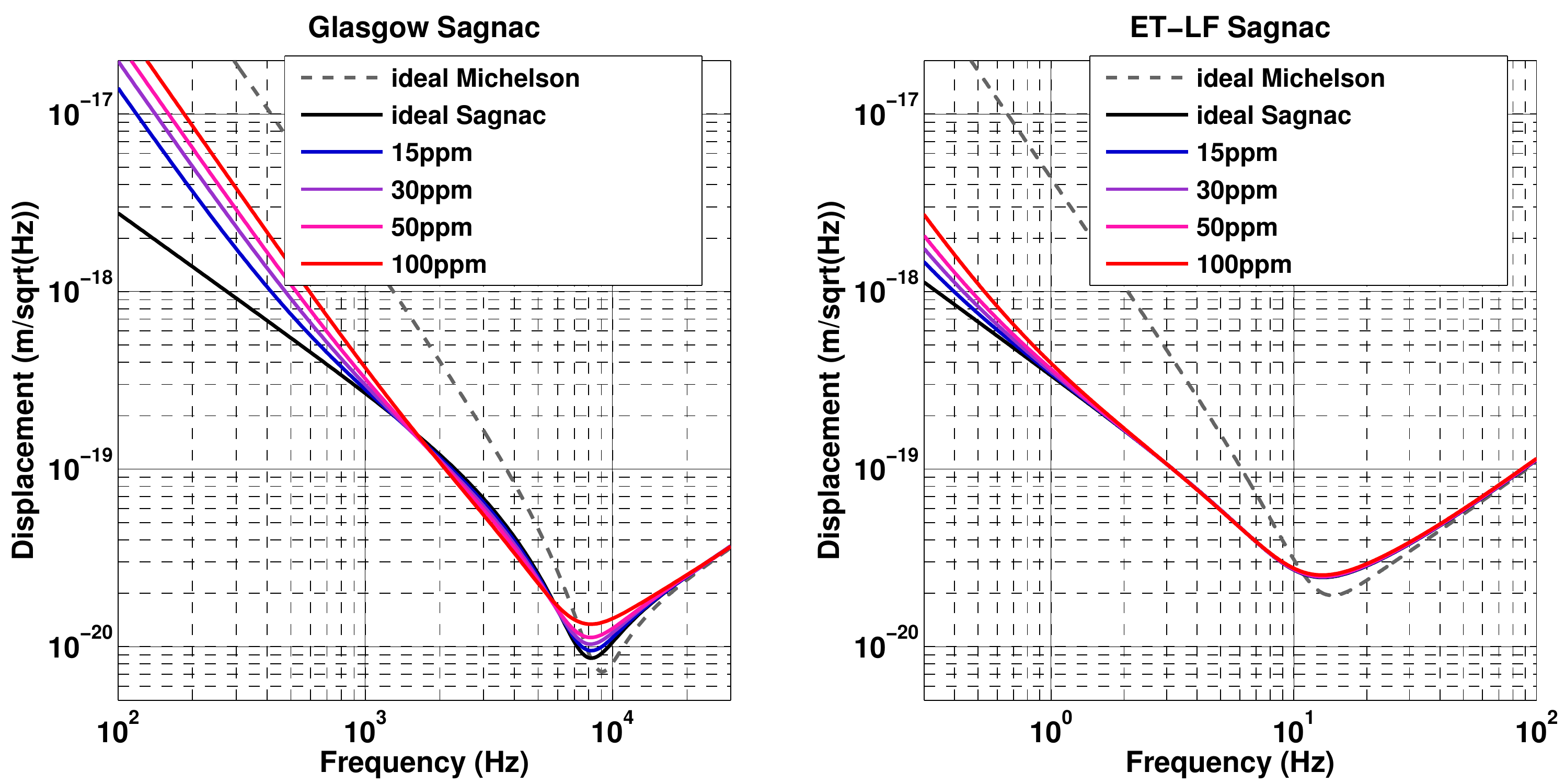}
  \caption{Quantum noise limited sensitivity of the Glasgow 
  Sagnac speed meter proof of concept experiment (left)
  and a low frequency ET Sagnac interferometer (right) for 
  symmetric losses in the two ring cavities in the arms. \SD{Ideal Michelson interferometer parameters match those of the corresponding ideal Sagnac interferometer, shown in the same plot.}}
  \label{fig:SSM_vs_ET_symmetric_loss}
\end{figure}

The parameters under consideration for the Einstein Telescope low
frequency interferometer (ET-LF) were primarily taken from the
most recent sensitivity study at the time of writing \cite{Hild2011}.
Since this design includes power recycling whereas the Glasgow
speed meter experiment does not, the input power for ET-LF
has been increased from $3.00$\,W to $45.73$\,W to account for
the lack of power recycling cavity gain. This change maintains
the intended circulating cavity power of $18$\,kW. Additionally,
to maintain the frequency at which the interferometer is most
sensitive, the transmissivity of the cavity input mirrors has been
altered from $7000$\,ppm to $10000$\,ppm. This recovers in our
model the frequency at which the ET-LF interferometer is intended
to be most sensitive. A list of parameters relevant to the model
is shown for our ET-LF Sagnac interferometer in the right hand
column of Table \ref{tab:etlf_params}.

Figure~\ref{fig:SSM_vs_ET_symmetric_loss} shows how symmetric
losses, i.e. losses that are identical in both ring  
cavities, degrade the quantum noise limited sensitivity of 
our two example configurations. The black traces represent 
perfectly balanced optical configurations with no losses in the 
interferometer arms. The remaining traces indicate symmetric 
losses in the range from 15\,ppm to 100\,ppm.\footnote{\SD{In real interferometers, 
the actual value of round-trip loss depends strongly on the length of the cavities. 
Longer cavities are known to be more lossy than the shorter ones (see \cite{2013_PhysRevD.88.022002_Evans_FCs,Miao14}). Here, however, 
we use the same value for both the short- and the long-base interferometers in order to make 
a fair comparison between them and make the effect of arm length on the impact of imperfections more profound.}} As has been 
described in the previous section, the main effect of the 
losses in the arm cavities shows up as an increased level of
quantum noise at low frequencies, which features a 1/$f^2$ slope. 
Overall, the loss-driven increase of the quantum noise limited 
sensitivity is much stronger for the Glasgow speed meter 
than it is for the ET-LF speed meter. This can be understood by
considering the fact that the Glasgow speed meter possesses arm cavity
finesse approximately 20 times higher than those of the ET-LF
Sagnac configuration. Despite similar round trip loss, the
total loss experienced in the short Glasgow speed meter arm cavities
is therefore about 20 times higher than for that of the low
frequency ET interferometer. 

It should be noted that the quantum noise with losses for the 
short Glasgow speed meter cannot be calculated accurately using
the approximation that the arm cavity round trip losses are small 
compared to the input mirror transmission. Doing so would strongly
underestimate the effect of the losses. It is therefore crucial
that all quantum noise calculations for the Glasgow speed meter
experiment fully account for losses (without relying on approximations),
as we have done in the analysis presented in this article.

\begin{figure}[h]
  \centering
  \includegraphics[width=\textwidth]{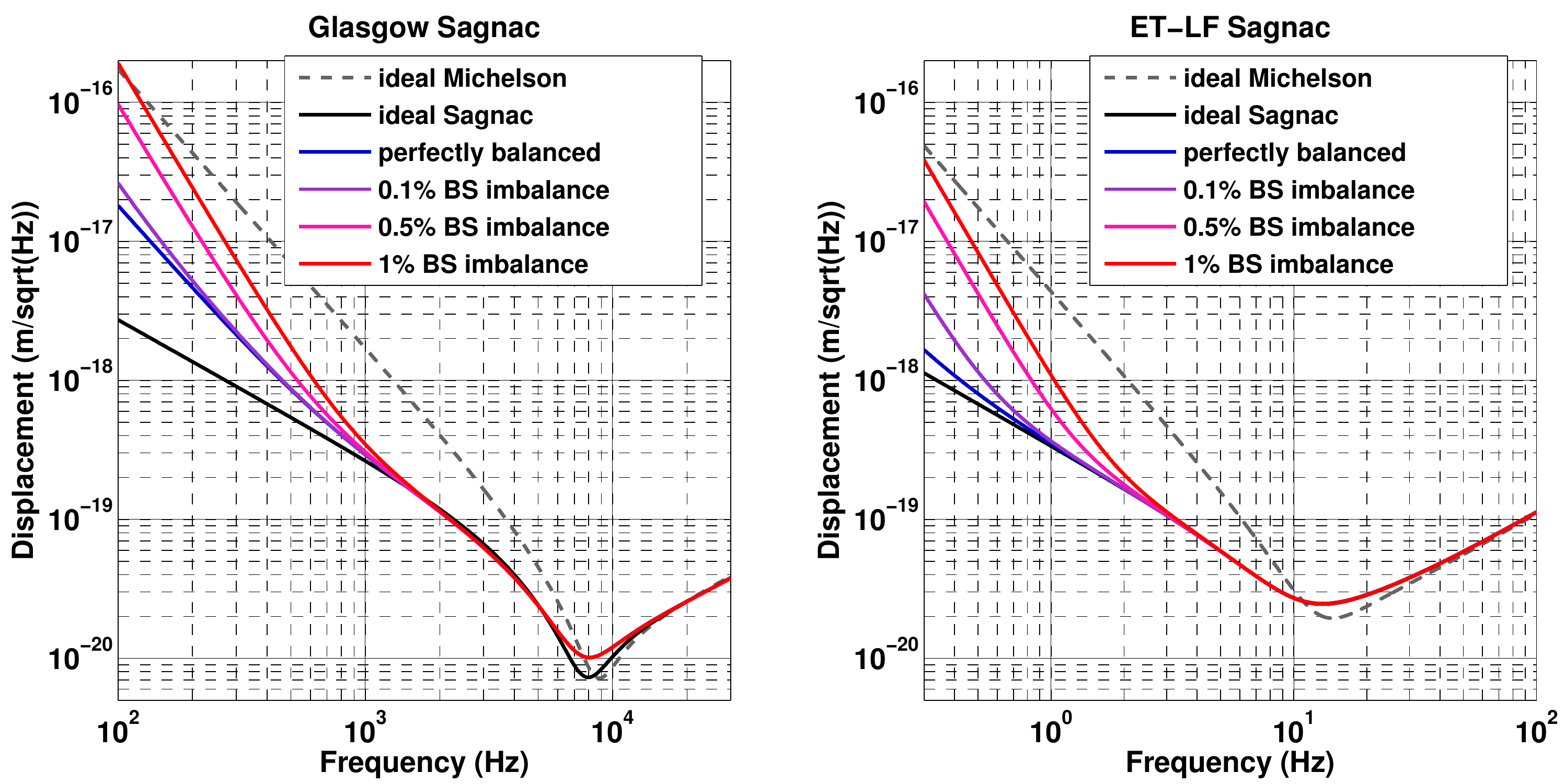}
  \caption{Quantum noise limited sensitivity of the Glasgow 
  Sagnac speed meter proof-of-concept experiment (left)
  and a low frequency ET Sagnac interferometer (right) for 
  an asymmetric beam splitter. \SD{Ideal Michelson interferometer parameters match those of the corresponding ideal Sagnac interferometer, shown in the same plot.} (Note that all traces apart from
   the ones labelled 'ideal', are calculated with symmetric
    arm cavity losses of 25\,ppm.)}
  \label{fig:SSM_vs_ET_bs_splitting_imbalance}
\end{figure}

Figure~\ref{fig:SSM_vs_ET_bs_splitting_imbalance} shows the 
influence of an imbalance in the reflection to transmission 
ratio of the main interferometer beam splitters. Please note 
that the coloured traces represent configurations with nominal 
arm cavity losses (i.e. 25\,ppm) and different levels 
of beam 
splitter imbalance, while for reference the black traces indicate
the case of no losses and perfectly balanced transmission and
reflection. For a beam splitter
imbalance of the order 0.1\,\% we find that the slope of the 
quantum noise at low frequencies approaches a 1/$f^3$ slope,
as was discussed and explained earlier in this article.

At first glance it might seem that the ET speed meter 
tends to be more susceptible to beam splitter imbalance
than the Glasgow speed meter (by comparing the separation of the red
and dark blue traces). However, in reality this difference only originates
from the fact that for a perfectly balanced system the quantum noise
of the Glasgow speed meter is already degraded much more from 
the 25\,ppm round trip loss than the quantum noise of the ET 
interferometer. If we compare the quantum noise with beam splitter
imbalance (blue traces) to the case of no losses combined with 
perfect beam splitter balance (black traces), then the overall quantum noise 
degradation looks similar for the two example configurations. This 
can be intuitively understood by considering that a beam splitter
imbalance causes a reduction in the cancellation of quantum noise, which 
is independent of the arm cavity finesse.   

\begin{figure}[h]
  \centering
  \includegraphics[width=\textwidth]{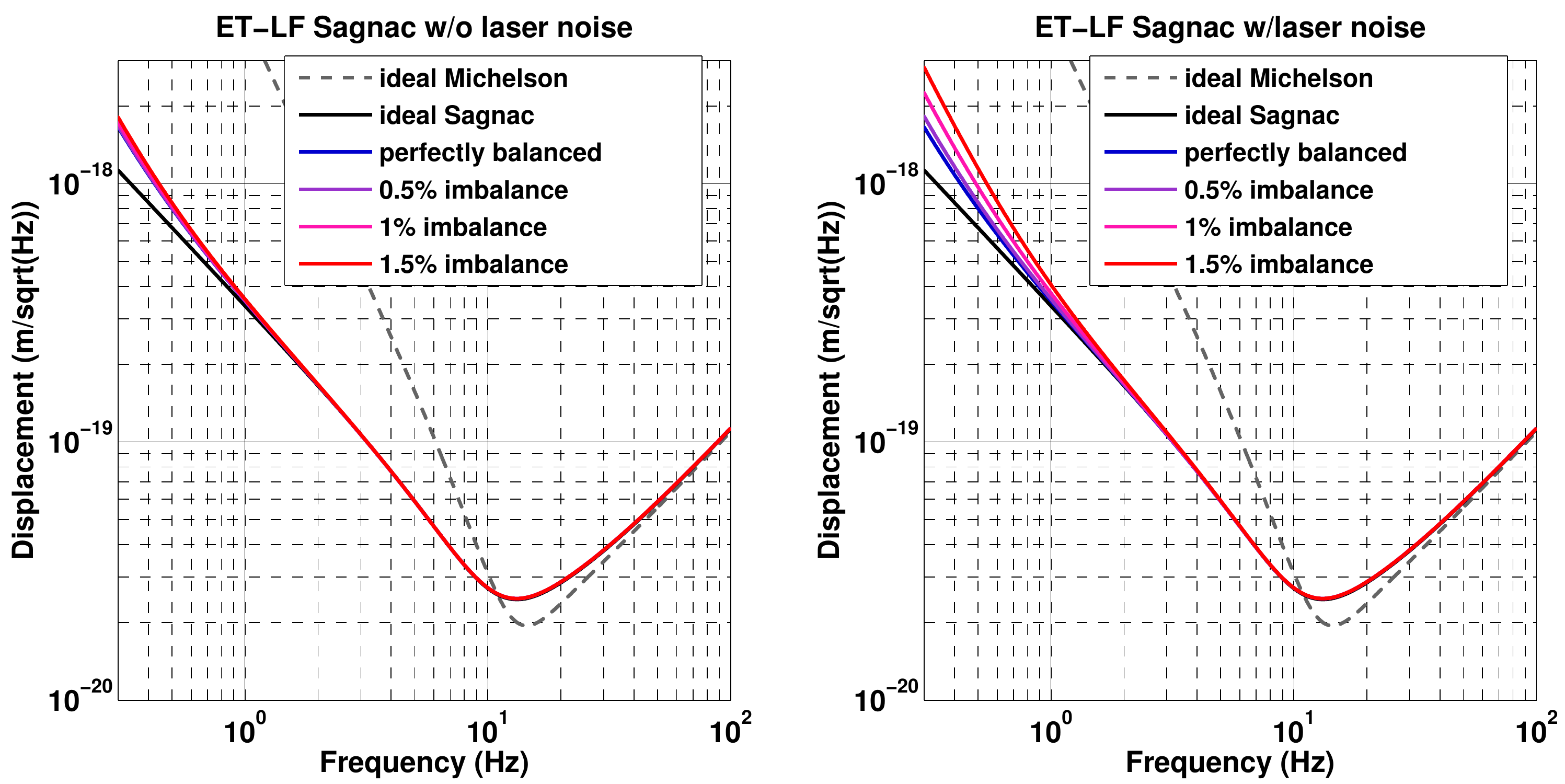}
  \caption{Quantum noise limited sensitivity of the ET Sagnac interferometer in case of asymmetric reflectivities of the ITMs. \textit{Left panel} shows the influence of this asymmetry when there is no excess laser noise and pump laser is considered ideal. \textit{Right panel} demonstrates the impact this asymmetry makes in presence of excess laser noise amounting to 10 times the vacuum level in power in both, the amplitude and the phase quadratures. \SD{Ideal Michelson interferometer parameters match those of the corresponding ideal Sagnac interferometer, shown in the same plot.} (Note that all traces apart from the ones labelled 'ideal', are calculated with symmetric arm cavity losses of 25\,ppm.) }
  \label{fig:SSM_vs_ET_ITM_transmissivity_imbalance}
\end{figure}

Finally, Figure~\ref{fig:SSM_vs_ET_ITM_transmissivity_imbalance}
illustrates the effect of imbalance of the reflectivities
of the two input mirrors combined with the effect of laser noise.
Both plots are based on the ET configuration with asymmetric
arm cavity input mirror reflectivities. However, the left plot
assumes an ideal laser, i.e. the laser output is
 limited by vacuum 
noise, while in the right hand plot the presence of excess 
noise of 10 times the vacuum is assumed be present in 
both quadratures on the laser. 
As can be seen from this comparison the excess laser noise 
significantly increases the effect of the imbalances in the 
interferometer configuration.\footnote{The effect shown here 
is even more profound for a BS imbalance.}   

\section{Summary}
\label{sec:summary}
In this article we have developed for the first time 
an analytical analysis that can accurately predict
the quantum noise limited sensitivity of Sagnac speed
meter interferometers featuring arm cavities. In particular,
our models do not reply on the common assumption that the 
arm cavity round trip loss is small compared to the arm 
cavity input mirror transmission. 

We have illustrated the results of our analysis by applying 
the model to two different speed meter configurations 
on very different length scales. We find that for 
the Glasgow speed meter proof-of-concept experiment,
symmetric arm cavity losses and beam splitter imbalance have 
the strongest influence on the achievable quantum noise level, 
while input mirror imbalances seem to be not too critical. 
In contrast, we find that for a 10\,km long ET Sagnac
interferometer the most significant quantum noise degradation
is caused by beam splitter imbalances, while symmetric losses
and input mirror imbalance play only a minor role.

\ack{We greatly appreciate all the help and illuminating discussions with A. Freise, D. Brown, H. Miao and S. Vyatchanin. We would also like to thank K. Strain for his constant support and constructive feedback.
The work described in this article is funded by the
European Research Council (ERC-2012-StG: 307245). We are grateful for support 
from Science and Technology Facilities Council (Grant Ref: ST/L000946/1), the Humboldt
Foundation, the International Max Planck Partnership (IMPP) and the ASPERA 
ET-R\&D project.}

\appendix

\section{Derivation of input-output relations for imperfect zero-area Sagnac interferometer.}\label{app_sec:I/O_relations_imp_sag}
In this section, we present a detailed derivation of I/O-relations for an imperfect Sagnac interferometer and derive an unabridged expression for the quantum noise spectral density. We start with the lossy arm cavity relations, then proceed to the imperfect, lossy beam splitter relations and, finally, derive the expressions for output fields of the entire Sagnac interferometer expressed in terms of the input fields. 

\subsection{Arm cavity input-output relations}\label{app_sec:I/O-rel_arms_full}
The general the I/O-relations of a lossy arm cavity of a Sagnac interferometer can be written as follows
\begin{multline}\label{app_eq:I/O-rels_arms_gen}
\vq{b}^{IJ} = \tq{T}^{IJ}_{\rm arm}\cdot\vq{a}^{IJ} + \tq{N}^{IJ}_{\rm arm}\cdot\vq{n}^{IJ} + \tq{T}_{\rm arm,\ r.p.}^{\bar{I}J}\cdot\vq{a}^{\bar{I}J} +\tq{N}_{\rm arm,\ r.p.}^{\bar{I}J}\cdot\vq{n}_{\bar{I}J} + \vs{R}^{IJ}_{\rm arm}\frac{x_J}{x_{\rm SQL}}\,.
\end{multline}
 To calculate radiation pressure contribution to the transfer matrices as well as to account for effects of cavity detuning on the mirrors' dynamics, we need to calculate the intracavity field as a function of the input fields as well:
\begin{align}\label{app_eq:intracav}
\vq{e}^{IJ} = \frac{1}{\sqrt{\tau}}\tq{L}_J(\Omega)\cdot\Bigl[\sqrt{\gamma^{J}_{\rm ITM}}\vq{a}^{IJ} + \sqrt{\gamma^{J}_{\rm loss}}\vq{n}^{IJ} \Bigr] + \frac{1}{2\sqrt{\gamma^{J}_{\rm ITM}\tau}} \vs{R}^{IJ}_{\rm arm}\dfrac{x_J}{x_{\rm SQL}}\,,
\end{align}
where again $J = E,N$ and $\tau = L/c$ is the light travel time between the arm cavity mirrors,
\begin{align}\label{app_eq:I/O_arms_TMs}
  \tq{T}^{IJ}_{\rm arm} &= 2\gamma^J_{\rm ITM} \mathbb{L}_J(\Omega) - \mathbb{I} +  \tq{T}_{\rm r.p.}^{IJ}\,, &
  \tq{N}^{IJ}_{\rm arm} &= 2\sqrt{\gamma^J_{\rm ITM}\gamma^J_{\rm loss}} \mathbb{L}_J(\Omega) + \tq{N}_{\rm r.p.}^{IJ}\,.
\end{align}
Here
\begin{gather}
  \mathbb{L}_J(\Omega) = \frac{1}{\mathcal{D}_{J}(\Omega)}
    \begin{bmatrix}
    \gamma^J_{\rm ITM}+\gamma^J_{\rm loss}-i\Omega & -\delta_J\\
     \delta_J & \gamma^J_{\rm ITM}+\gamma^J_{\rm loss}-i\Omega
    \end{bmatrix}\,, \label{FP_bbL} \\
  \mathcal{D}_J(\Omega) =  (\gamma^J_{\rm ITM}+\gamma^J_{\rm loss}-i\Omega)^2 + \delta_J^2 \,,
\end{gather}
\begin{equation}
    \vs{R}^{IJ}_{\rm arm}(\Omega)= \sqrt{\frac{16\omega_pP^{IJ}_c \gamma^{J}_{\rm ITM}}{\mu^J_{\rm arm}\Omega^2 L c}}\mathbb{L}(\Omega)\cdot
    \begin{bmatrix}
      -\sin\Phi_{IJ}\\
      \cos\Phi_{IJ}
    \end{bmatrix} = 
    \sqrt{\frac{4\Theta^{IJ}\gamma^{J}_{\rm ITM}}{\Omega^2}}\mathbb{L}(\Omega)\cdot
    \begin{bmatrix}
      0\\
      1
    \end{bmatrix} 
    \,.
\end{equation} 
In the equantions above $\Phi_{IJ}$ stands for phases, the $IJ$-beam field has at the ITM of the $J$-th arm cavity. Its choice is arbitrary and depends on the chosen carrier field reference phase, so we can always set it to zero, as it is done in the second equation. $\Theta^{IJ} = 4\omega_p P^{IJ}_c/(\mu_{\rm arm}^J cL)$ is the normalised power circulating in the $I$-th arm in the $J$-th beam, $\mu_{\rm arm}^J = 2M^J_{\rm ITM}M^J_{\rm ETM}/(M^J_{\rm ITM}+2M^J_{\rm ETM})$ is an effective mass of the $J$-th arm with $M^J_{\rm ITM}$ and $2M^J_{\rm ETM}$ being the masses of ITM and ETM (note that there are 2 of them in each cavity),  $\delta_J = \omega_J - \omega_p$ is the $J$-th cavity resonance frequency, $\omega_J$, detuning from the pump laser frequency $\omega_p$, and the cavity half-bandwidths due to ITM power transmissivity, $T^J_{\rm ITM}$, and due to loss, $T^J_{\rm loss}$, read 
\begin{align}
\gamma^{J}_{\rm ITM} &= \dfrac{cT^J_{\rm ITM}}{4L}\,, & \gamma^{J}_{\rm loss} &= \dfrac{cT^J_{\rm loss}}{4L}\,.
\end{align}

To calculate the radiation pressure contribution we need to know how the mirrors move under the radiation pressure force from both beams. Writing down the equations of motion for each mirror and then combining them in the effective arm degree of freedom $x_J$, one can get the following expression for the latter in the frequency domain:
\begin{align}\label{app_eq:x_arm}
x_J &= x_J^{\rm signal}+\hat{x}_J^{r.p.}= x_J^{\rm signal}+\chi^J(\Omega) [\hat{F}^{IJ} + \hat{F}^{\bar{I}J}]\,, \mbox{where}\ \chi^J(\Omega) = -\frac{1}{\mu_{\rm arm}^J\Omega^2}\,,
\end{align}
where we assumed the dynamics of the arm to be that of a free mass with an effective mass $\mu_{\rm arm}^J$. In principle, it is always possible to introduce more complicated dynamics into our model by changing the shape of the mechanical susceptibility functions $\chi^J(\Omega)$. The radiation pressure forces created by each beam read:
\begin{multline}\label{app_eq:FRP_arm_full}
\hat{F}^{IJ} = 2 \frac{\hbar \omega_p}{c} (\vs{E}^{IJ})^{\rm T} \cdot \vq{e}_{IJ} = \sqrt{\dfrac{8\hbar\omega_pP^{IJ}_c}{c^2}}\begin{bmatrix}
      1\\
      0
    \end{bmatrix}^{\rm T}\cdot\vq{e}_{IJ} = \hat{F}^{IJ}_{\rm r.p.}-K^{IJ}_{\rm arm}(\Omega)x_J =\\
    =\sqrt{2\hbar\mu^J_{\rm arm}\Theta^{IJ}}\begin{bmatrix}
      1\\
      0
    \end{bmatrix}^{\rm T}\cdot \tq{L}_J(\Omega)\cdot\Bigl[\sqrt{\gamma^{J}_{\rm ITM}}\vq{a}^{IJ} + \sqrt{\gamma^{J}_{\rm loss}}\vq{n}^{IJ} \Bigr] - \dfrac{\mu^J_{\rm arm}\Theta^{IJ}\delta^J}{\mathcal{D}_J(\Omega)}x_J\,.
\end{multline}
Here the first term, $\hat{F}^{IJ}_{\rm r.p.}$, is the pure fluctuational force, and the last term, $K^{IJ}_{\rm arm}(\Omega)x_J$, is the dynamical back-action term with $K^{IJ}_{\rm arm}(\Omega)$ an optical rigidity, which is only relevant for non-zero arm detuning $\delta^J$.

Then we substitute the expression \eqref{app_eq:FRP_arm_full} into \eqref{app_eq:x_arm} and get the new equation for the cavity mirrors dynamics:
\begin{align*}
x_J &= x_J^{\rm signal}+\chi^J(\Omega) [\hat{F}^{IJ}_{\rm r.p.} + \hat{F}^{\bar{I}J}_{\rm rp} - (K^{IJ}_{\rm arm}+K^{\bar{I}J}_{\rm arm})x_J]\,, 
\end{align*}
which can be resolved in $x_J$ to give:
\begin{align}\label{app_eq:x_arm_mod}
x_J &= x_J^{\rm signal}+\chi^J_{\rm new}(\Omega) [\hat{F}^{IJ}_{\rm r.p.} + \hat{F}^{\bar{I}J}_{\rm rp}]\,, 
\end{align}
where the new modified mechanical susceptibility reads:
\begin{align}
\chi^J_{\rm new}(\Omega) = \dfrac{\chi^J(\Omega)}{1+\chi^J(\Omega)(K^{IJ}_{\rm arm}(\Omega)+K^{\bar{I}J}_{\rm arm}(\Omega))}\,.
\end{align}
Note that for cavities tuned to resonance, $\chi^J_{\rm new}(\Omega)=\chi^J(\Omega)$.

The expressions for $\tq{T}_{\rm r.p.}^{IJ}$ and $\tq{N}_{\rm r.p.}^{IJ}$ are obtained by substituting \eqref{app_eq:x_arm_mod} into the following formula, representing the back-action induced contribution to the output field:
\begin{align}
\Delta\vq{b}^{IJ}_{\rm r.p.} = \vs{R}^{IJ}_{\rm arm}\frac{x_J-x_J^{\rm signal}}{x_{\rm SQL}}\,,
\end{align}
and collecting the coefficients in front of the corresponding light field. Thereby we arrive at the following expressions:
\begin{align}
 \tq{T}^{IJ}_{\rm r.p.} &= 
 2\mu^J_{\rm arm}\chi^J_{\rm new}\Theta^{IJ}\gamma^J_{\rm ITM}\tq{L}_J(\Omega)\cdot
 \begin{bmatrix}
 0 & 0\\
 1 & 0
 \end{bmatrix}
 \cdot\tq{L}_J(\Omega)\,, \label{app_eq:T_RP_mat_gen}\\
  \tq{N}^{IJ}_{\rm r.p.} &= 2\mu^J_{\rm arm}\chi^J_{\rm new}\Theta^{IJ}\sqrt{\gamma^J_{\rm ITM}\gamma^J_{\rm loss}}\tq{L}_J(\Omega)\cdot
 \begin{bmatrix}
 0 & 0\\
 1 & 0
 \end{bmatrix}
 \cdot\tq{L}_J(\Omega)\,.\label{app_eq:N_RP_mat_gen}
\end{align}

The two fields leaving the interferometer and mixing at the beam splitter are $\vq{b}^{LN}$ and $\vq{b}^{RE}$. They can be expressed in terms of the input fields, $\vq{a}^{RN}$ and $\vq{a}^{LE}$, as well as of noise fields $\vq{n}^{IJ}$ using continuity conditions:
\begin{align}
\vq{a}^{LN} &= \vq{b}^{LE}\,, & \vq{a}^{RE} &= \vq{b}^{RN}\,.
\end{align}
Then the general expression for each arm's  I/O-relations read:
\begin{subequations}\label{app_eq:I/O-rels_arms_resolved}
\begin{align}
\vq{b}^{LN} &= \tq{T}^{LN}_{\rm arm}\left[(\tq{I}-\tq{T}^{RE}_{\rm r.p.}\tq{T}^{LN}_{\rm r.p.})^{-1}\tq{T}^{RE}_{\rm r.p.}\vq{f}^{RN}+(\tq{I}-\tq{T}^{RE}_{\rm r.p.}\tq{T}^{LN}_{\rm r.p.})^{-1}\vq{f}^{LE}\right]+\vq{f}^{LN}\,,\label{app_eq:arm_bLN} \\
\vq{b}^{RE} &= \tq{T}^{RE}_{\rm arm}\left[(\tq{I}-\tq{T}^{LN}_{\rm r.p.}\tq{T}^{RE}_{\rm r.p.})^{-1}\vq{f}^{RN}+(\tq{I}-\tq{T}^{LN}_{\rm r.p.}\tq{T}^{RE}_{\rm r.p.})^{-1}\tq{T}^{LN}_{\rm r.p.}\vq{f}^{LE}\right] + \vq{f}^{RE}\,,\label{app_eq:arm_bRE}
\end{align}
where
\begin{align}
\vq{f}^{LN} &= \tq{T}^{RN}_{\rm r.p.}\vq{a}^{RN} + \tq{N}^{LN}_{\rm arm}\vq{n}^{LN} + \tq{N}^{RN}_{\rm r.p.}\vq{n}^{RN} + \vs{R}^{LN}_{\rm arm}\frac{x_N}{x_{\rm SQL}}\,,\\
\vq{f}^{RN} &=\tq{T}^{RN}_{\rm arm}\vq{a}^{RN} +  \tq{N}^{RN}_{\rm arm}\vq{n}^{RN} + \tq{N}^{LN}_{\rm r.p.}\vq{n}^{LN} + \vs{R}^{RN}_{\rm arm}\frac{x_N}{x_{\rm SQL}}\,,\\
\vq{f}^{LE} &=\tq{T}^{LE}_{\rm arm}\vq{a}^{LE} +  \tq{N}^{LE}_{\rm arm}\vq{n}^{LE} + \tq{N}^{RE}_{\rm r.p.}\vq{n}^{RE} + \vs{R}^{LE}_{\rm arm}\frac{x_E}{x_{\rm SQL}}\,,\\
\vq{f}^{RE} &= \tq{T}^{LE}_{\rm r.p.}\vq{a}^{LE} +  \tq{N}^{RE}_{\rm arm}\vq{n}^{RE} + \tq{N}^{LE}_{\rm r.p.}\vq{n}^{LE} + \vs{R}^{RE}_{\rm arm}\frac{x_E}{x_{\rm SQL}}\,.
\end{align}
\end{subequations}

\paragraph{Special case of resonant arms:} These bulky relations become significantly simpler as the arm cavities are set to resonance, \textit{i.e.} for $\delta^J = 0$. Then the radiation pressure matrices defined in \eqref{app_eq:T_RP_mat_gen} and \eqref{app_eq:N_RP_mat_gen} take the much simpler form:
\begin{align}
 \tq{T}^{IJ}_{\rm r.p.} &= 
e^{2i\beta^J_{\rm arm}}
 \begin{bmatrix}
 0 & 0\\
 -\mathcal{K}^{IJ}_{\rm arm} & 0
 \end{bmatrix}\,,  &
  \tq{N}^{IJ}_{\rm r.p.} &=  \sqrt{\frac{\gamma^J_{\rm loss}}{\gamma^J_{\rm ITM}}} \tq{T}^{IJ}_{\rm r.p.}\,, & \vs{R}^{IJ}_{\rm arm} &= \sqrt{2\mathcal{K}^{IJ}_{\rm arm}}e^{i\beta^J_{\rm arm}}
  \begin{bmatrix}
  0\\
  1
  \end{bmatrix}\,,\label{app_eq:N_RP_mat_res}
\end{align}
where optomechanical coupling factor of a lossy arm is defined as:
\begin{equation}
\mathcal{K}^{IJ}_{\rm arm} = \dfrac{\Theta^{IJ}\gamma^J_{\rm ITM}}{\Omega^2[(\gamma^J_{\rm ITM}+\gamma^J_{\rm loss})^2+\Omega^2]}\,,\quad\beta^J_{\rm arm} = \arctan\frac{\Omega}{\gamma^J_{\rm ITM}+\gamma^J_{\rm loss}}\,.
\end{equation}
In this particular case, the radiation pressure matrices $\tq{T}^{IJ}_{\rm r.p.}$ and $\tq{N}^{IJ}_{\rm r.p.}$ are orthogonal to each other, meaning that any product of them, irrespective of what value the indices $I,J$ have, is zero.  Transfer matrices  \eqref{app_eq:I/O_arms_TMs} become:
\begin{align}\label{app_eq:I/O_arms_TMs_res}
  \tq{T}^{IJ}_{\rm arm} &= e^{2 i\beta_{\rm arm}^J}
  \begin{bmatrix}
  \mathcal{T}_{\rm arm}^J& 0\\
  -\mathcal{K}^{IJ}_{\rm arm} & \mathcal{T}_{\rm arm}^J
  \end{bmatrix}\,, &
  \tq{N}^{IJ}_{\rm arm} &= 
 \sqrt{\frac{\gamma^J_{\rm loss}}{\gamma^J_{\rm ITM}}} e^{2 i\beta_{\rm arm}^J}
  \begin{bmatrix}
  \mathcal{N}_{\rm arm}^J & 0\\
  -\mathcal{K}^{IJ}_{\rm arm} & \mathcal{N}_{\rm arm}^J
  \end{bmatrix}\,,
\end{align}
where
\begin{align*}
 \mathcal{T}_{\rm arm}^J(\Omega) &= \dfrac{\gamma^J_{\rm ITM}-\gamma^J_{\rm loss}+i\Omega}{\gamma^J_{\rm ITM}+\gamma^J_{\rm loss}+i\Omega}\,, & \mathcal{N}_{\rm arm}^J(\Omega) &= \dfrac{2\gamma^J_{\rm ITM}}{\gamma^J_{\rm ITM}+\gamma^J_{\rm loss}+i\Omega}\,.
\end{align*}
This simplifies the I/O-relations \eqref{app_eq:I/O-rels_arms_resolved} substantially:
\begin{subequations}\label{app_eq:I/O-rels_arms_res}
\begin{align}
\vq{b}^{LN} &= \tq{T}^{LN}_{\rm arm}\left[\tq{T}^{RE}_{\rm r.p.}\vq{f}^{RN}+\vq{f}^{LE}\right]+\vq{f}^{LN}\,, \\
\vq{b}^{RE} &= \tq{T}^{RE}_{\rm arm}\left[\vq{f}^{RN}+\tq{T}^{LN}_{\rm r.p.}\vq{f}^{LE}\right] + \vq{f}^{RE}\,,
\end{align}
\end{subequations}

These simplified expressions can be used to estimate the influence of different asymmetries on the Sagnac interferometer sensitivity. To make the final step in the calculation of the spectral density, we need to refer to the beam splitter relations, which is presented in the next subsection:

\subsection{Beam splitter input/output relations}
The input and output fields of the beam splitter are shown in Fig.~\ref{fig:BS}. The corresponding input-output relations read:
\begin{subequations}\label{app_eq:I/O-rels_BS_full}
\begin{align}
\vq{o} &= \sqrt{1-\epsilon_{\rm BS}} (-\sqrt{R_{\rm BS}} \vq{b}^{RE} + \sqrt{T_{\rm BS}} \vq{b}^{LN}) + \sqrt{\epsilon_{\rm BS}} \vq{m}_o\, , \label{app_eq:bs_o}\\
\vq{q} &= \sqrt{1-\epsilon_{\rm BS}} (\sqrt{T_{\rm BS}} \vq{b}^{RE} + \sqrt{R_{\rm BS}} \vq{b}^{LN}) + \sqrt{\epsilon_{\rm BS}} \vq{m}_p\, , \label{app_eq:bs_d}\\
\vq{a}^{RN} &= \sqrt{T_{\rm BS}} (\sqrt{1-\epsilon_{\rm BS}}\vq{i} + \sqrt{\epsilon_{\rm BS}} \vq{m}_i) + \sqrt{R_{\rm BS}} (\sqrt{1-\epsilon_{\rm BS}}\vq{p} + \sqrt{\epsilon_{\rm BS}} \vq{m}_p) \, ,\label{app_eq:bs_aRN}\\ 
\vq{a}^{LE} &= -\sqrt{R_{\rm BS}} (\sqrt{1-\epsilon_{\rm BS}}\vq{i} + \sqrt{\epsilon_{\rm BS}} \vq{m}_i) + \sqrt{T_{\rm BS}} (\sqrt{1-\epsilon_{\rm BS}}\vq{p} + \sqrt{\epsilon_{\rm BS}} \vq{m}_p)\,. \label{app_eq:bs_aLE}
\end{align}
\end{subequations}
We introduced a BS asymmetry offset, $\alpha_{\rm BS}\ll1$, in Eq.~\eqref{eq:asym_BS_offset}. Losses at the beam splitter are accounted for by introducing the loss factor $\epsilon_{\rm BS}\ll1$ and corresponding vacuum fields, $\vq{m}_{i,p}$.  
Substituting equations Eqs.~\eqref{app_eq:bs_aRN}, \eqref{app_eq:bs_aLE} into Eqs.~\eqref{app_eq:arm_bLN}, \eqref{app_eq:arm_bRE} and substituting the result into Eq.~\eqref{app_eq:bs_o}, we finally get the full interferometer I/O relations: 
\begin{align}
 \vq{o} &= \tq{T}^i_{\rm sag}\cdot\vq{i} + \tq{T}^p_{\rm sag}\cdot\vq{p} + \sum\limits_{\substack{I=L,R\\ J=N,E}}\tq{N}^{IJ}_{\rm sag}\cdot\vq{n}_{IJ} \sum\limits_{k=i,p}\tq{M}^{k}_{\rm sag}\cdot\vq{m}_{k} + \vb{R}^{+}_{\rm sag}x_++\vb{R}^{-}_{\rm sag}x_-\,.
\end{align}
Here $\tq{M}^{i,p}_{\rm sag}$ stand for transfer matrices for additional noise associated with the BS loss.

Collecting the terms in front of corresponding vacuum fields and mechanical displacement terms, one can get the unabridged expressions for transfer matrices and represent the I/O  relations. These expressions are rather cumbersome and opaque, though straightforward to derive, so we omit them here. The quantum noise power spectral density can be then calculated using the general rule \eqref{eq:SpDens_x_loss}, which yields:
 \begin{multline}\label{app_eq:SpDens_x_asSag}
  S^x(\Omega) = \frac{x^2_{\rm SQL}}{|\vs[T]{H}_\zeta\cdot\vs{R}^-_{\rm sag}|^2}\Bigl\{\vs[T]{H}_\zeta\cdot[\tq{T}^i_{\rm sag}\cdot\tq{S}^{in}_{i}\cdot(\tq{T}^i_{\rm sag})^\dag+\tq{T}^p_{\rm sag}\cdot(\tq{T}^p_{\rm sag})^\dag]\cdot\vs{H}_\zeta + 
 \\ + \sum\limits_{\substack{I=L,R\\ J=N,E}}\vs[T]{H}_\zeta\cdot\tq{N}^{IJ}_{\rm sag}\cdot(\tq{N}^{IJ}_{\rm sag})^\dag\cdot\vs{H}_\zeta+\sum\limits_{k=i,p}\vs[T]{H}_\zeta\cdot\tq{M}^{k}_{\rm sag}\cdot(\tq{M}^{k}_{\rm sag})^\dag\cdot\vs{H}_\zeta\Bigr\}\,.
\end{multline}
Here we normalised quantum noise to the \textit{dARM} signal, as indicated by the denominator, where $\vs{R}^-_{\rm sag}$ stands for the interferometer response function to differential motion of the mirrors.

\section{Laser noise in asymmetric Sagnac interferometer.}
\begin{figure}[h]
\begin{center}
	\includegraphics[width=.9\textwidth]{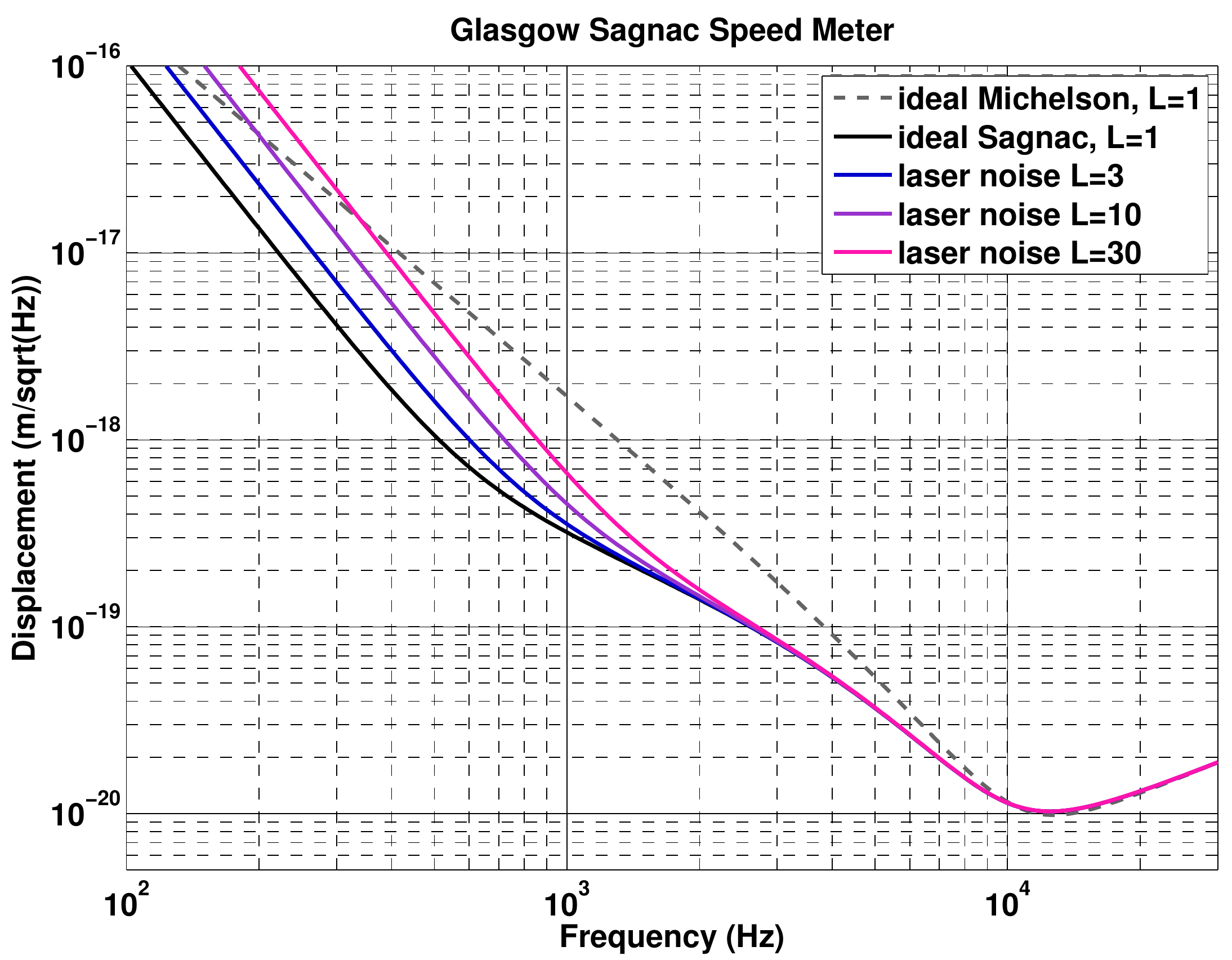}
\caption{Spectral density plots for a table-top Sagnac interferometer with parameters given in Table.~\ref{tab:etlf_params}. The influence of laser noise in the presence of 1\% asymmetry of the beam splitter
for different levels of laser noise. We assume fluctuations of two quadratures of laser light independent and having the same spectral density $L$ which takes the values $L=1,\, 3\,, 10\,, 30$ times the vacuum level. \SD{Ideal Michelson interferometer parameters match those of the corresponding ideal Sagnac interferometer, shown in the same plot.}}
\label{app_fig:AS_Sag_SpDens_LasNoise}
\end{center}
\end{figure}

The main implication an asymmetry of the interferometer has in regards to the quantum noise is the leakage of laser noise to the output port. Our approach allows to account for this effect assuming a simple model of laser noise as an excess fluctuation on top of the quantum uncertainties of the input laser light. If we assume that the amplitude and phase fluctuations of the carrier light are uncorrelated and characterised by spectral densities $L_c>1$ and $L_s>1$, respectively, then the input state of the common mode light field $\vq{p}$ reads:
\begin{equation}
\tq{S}^{in}_p = 
\begin{bmatrix}
L_c & 0\\
0 & L_s
\end{bmatrix}\,,
\end{equation}
and the general quantum noise spectral density formula \eqref{app_eq:SpDens_x_asSag} shall be slightly modified to:
 \begin{multline}\label{app_eq:SpDens_x_asSag+laser_noise}
  S^x(\Omega) = \frac{x^2_{\rm SQL}}{|\vs[T]{H}_\zeta\cdot\vs{R}^-_{\rm sag}|^2}\Bigl\{\vs[T]{H}_\zeta\cdot[\tq{T}^i_{\rm sag}\cdot\tq{S}^{in}_{i}\cdot(\tq{T}^i_{\rm sag})^\dag+\tq{T}^p_{\rm sag}\cdot\tq{S}^{in}_p\cdot(\tq{T}^p_{\rm sag})^\dag]\cdot\vs{H}_\zeta + 
 \\ + \sum\limits_{\substack{I=L,R\\ J=N,E}}\vs[T]{H}_\zeta\cdot\tq{N}^{IJ}_{\rm sag}\cdot(\tq{N}^{IJ}_{\rm sag})^\dag\cdot\vs{H}_\zeta+\sum\limits_{k=i,p}\vs[T]{H}_\zeta\cdot\tq{M}^{k}_{\rm sag}\cdot(\tq{M}^{k}_{\rm sag})^\dag\cdot\vs{H}_\zeta\Bigr\}\,.
\end{multline}
The effect that such laser noise has on the quantum noise sensitivity is shown in Fig.~\ref{app_fig:AS_Sag_SpDens_LasNoise}. \SD{The chosen span of $L$ values starts at the shot noise level of $L=1$, which for the 1.7W laser to be used in the Glasgow prototype  Sagnac interferometer corresponds to the relative intensity noise (RIN) amplitude spectral density (ASD) of $4.7 \times 10^{-10}\ \mathrm{Hz}^{-1/2}$. The upper value of $L=30$ corresponds to the level of RIN available for the same 1.7W laser with reasonable intensity pre-stabilisation, \textit{i.e.} to the RIN ASD of $\sim 1.4 \times 10^{-8}\ \mathrm{Hz}^{-1/2}$.}

\section*{References}
%
\providecommand{\newblock}{}

\end{document}